\documentclass[aps,twocolumn,superscriptaddress,showpacs,floatfix,longbibliography,nobibnotes]{revtex4-1}
\usepackage{amsmath,amssymb,amsfonts,bm,bbm}
\usepackage{color}
\usepackage[pdftex]{hyperref,graphicx}

\newcommand{\ket}[1]{| {#1} \rangle}
\newcommand{\expect}[1]{\langle {#1} \rangle}

\begin{document}
\title{Theory of coherent phase modes in insulating Josephson junction chains }

\author{Huan-Kuang Wu}
\affiliation{Department of Physics, Condensed Matter Theory Center and Joint Quantum Institute, University of Maryland, College Park, MD 20742, USA}

\author{Jay D. Sau}
\affiliation{Department of Physics, Condensed Matter Theory Center and Joint Quantum Institute, University of Maryland, College Park, MD 20742, USA}

\date{\today\\}

\begin{abstract}
Recent microwave reflection measurements of Josephson junction chains have suggested the presence of nearly coherent collective charge oscillations deep in the insulating phase. 
Here we develop a qualitative understanding of such coherent charge modes by studying the local dynamical 
polarizability of the insulating phase of a finite length sine-Gordon model. 
By considering parameters near the non-interacting fermion limit where the charge operator dominantly couples to soliton-antisoliton pairs of the sine-Gordon model, we find that the local dynamical polarizability shows an array of sharp peaks 
in frequency representing coherent phase oscillations on top of an incoherent background. 
The strength of the coherent peaks relative to the incoherent background 
 increases as a power law in frequency as well as exponentially as the Luttinger parameter approaches a critical value.
 The dynamical polarizability also clearly shows the insulating gap. 
We then compare the results in the high frequency limit to a perturbative estimate of phase-slip-induced decay of 
plasmons in the Josephson junction chain.
\end{abstract}

\maketitle

\section{Introduction} The quantum dynamics of many-body systems  been at the crux of recent 
conceptual developments such as the phenomena of many-body localization~\cite{mblreview} and AdS-CFT correspondence~\cite{adscft}. This recent interest represents an attempt to extend our knowledge beyond the understanding of static quantum 
phenomena in terms of quantum field theory using the renormalization group~\cite{sachdevbook}. It is becoming clearer that such 
traditional approaches, such as analytic continuation of imaginary time correlators~\cite{sachdevbook},
 are insufficient to discuss quantum dynamical phenomena, which are becoming more 
accessible in experiments. One test-bed for understanding such dynamical quantum phenomena is the 
study of conductance at the superconductor-insulator critical point. Despite the static appearance of dc conductance, it is technically defined as a limit of vanishingly small frequency.
The theory in two dimensions leads to a 
prediction of interesting consequences of the particle-vortex duality such as universal conductance~\cite{fisher}
that are borne out by experiments~\cite{goldman,kapitulnik,hebard}. Microwave 
measurements of the ac conductivity have revealed intriguing signatures of 
residual superfluid stiffness even in the insulating phase~\cite{pratap}.
The superfluid-insulator 
transition in the Bose-Hubbard model~\cite{sachdevbook,greiner}
 also turned out to be quite 
interesting in the context of ultra-cold atoms where an under-damped Higgs mode~\cite{podolsky} was observed~\cite{bloch}.

A conceptually simpler context where the superfluid-insulator transition (SIT) was predicted to occur~\cite{doniach} is that of one dimensional chains of Josephson junctions (JJ's). Such a system, in the 
vicinity of the SIT quantum critical point, is described by the sine-Gordon Hamiltonian 
~\cite{giamarchibook} (rescaled to fit the convention for SIT~\cite{mirlininsulator}) 
of the form 
\begin{align}
&H=\int dx\frac{v_c}{2}[\pi K j^2+(\pi K)^{-1}\partial_x\phi^2]+g\cos{2\phi},\label{HGiam}
\end{align} 
where $\partial_x\phi\propto \rho(x)$, which is the density of Cooper pairs in the system and $j$ 
is the current density operator that is  canonically conjugate to $\phi$ and $v_c$ is the speed of the superfluid phase oscillations (also 
equivalently the charge velocity). 
Since the microscopic origin of the charging energy is from a combination of junction and ground capacitance, 
as detailed in Appendix A and B, the 
microscopic Hamiltonian of the JJ chain can be quantitatively mapped to the sine-Gordon Hamiltonian Eq.~\ref{HGiam} 
over a large parameter range.
Note that at $g=0$, the superconducting phase variable (defined as $\theta\propto\int dx j(x)$) is algebraically ordered allowing 
us to define superfluid phase oscillations. Such oscillations may also be thought of as plasma oscillations of the charge 
density $\rho(x)$ accompanied by coherent oscillation of the current density $j(x)$.
The parameter $g$ in Eq.~\ref{HGiam} is related to the amplitude of 
phase slips in the chain that would manifest as decay of supercurrent~\cite{matveev,rastelli} and $K$ is the Luttinger parameter, which manifests as the inverse of impedance of the chain and 
determines the transport of the chain at weak values of $g$. 
\begin{figure}
\centering
	\includegraphics[width=0.5\textwidth,angle=0]{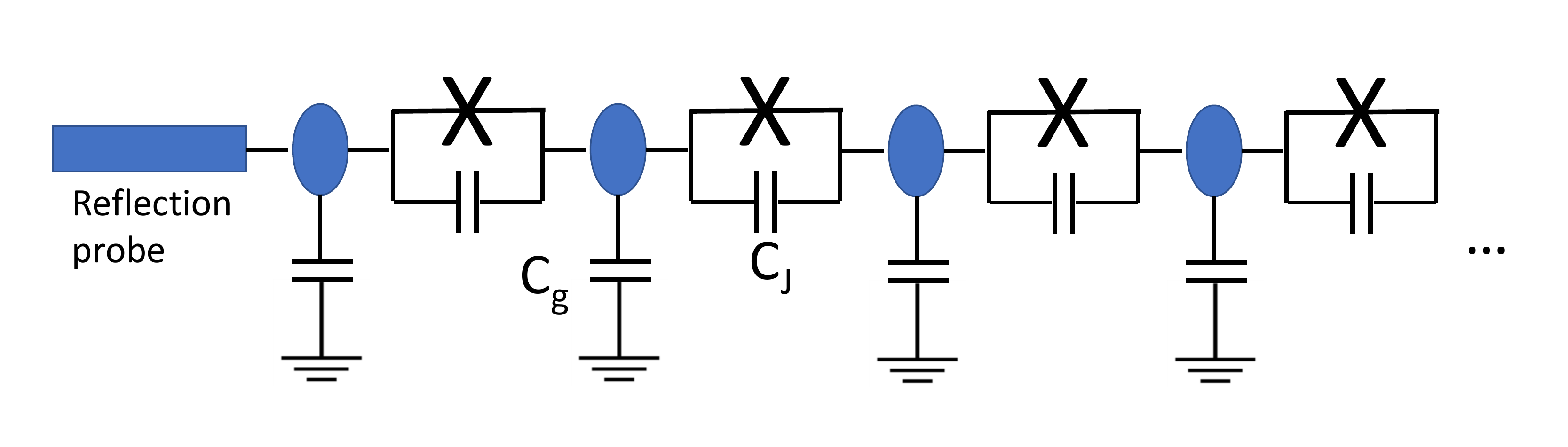}
	\caption{Absorption  of a weakly coupled transmission probe on the left measures the   dynamical conductivity  $\sigma_R(\omega)=\omega\chi(\omega)$ at the end of a JJ chain.
The JJ chain is composed of an array of islands with ground capacitance $C_g$ coupled by JJ's. The JJ's 
have a capacitance $C_J$ in addition to a Josephson coupling with strength $E_J$.
} 
	\label{Fig0}
\end{figure}
Specifically, the Hamiltonian $H$ would describe a near-critical insulating phase for the clean JJ chain at
 long wave-lengths for $K<2$ (or $3/2$ in the disordered case)~\cite{giamarchibook,mirlininsulator}.
 This transition was observed in SQUID arrays, where the effective 
Josephson coupling, which controls $K$ and $g$, are tuned across the transition~\cite{haviland}.

While the dc transport properties of the superconductor-insulator transition in JJ chains had been 
studied previously~\cite{haviland}, the dynamical properties have only recently begun to be explored~\cite{duty,vlad}.
One such study, which considered the microwave response (schematic shown in Fig.~\ref{Fig0}) of such JJ chains~\cite{vlad}, have revealed coherent 
oscillations associated with superfluid phase coherence deep in the insulating 
phase of such JJ chains, where the superfluid phase has been predicted to be 
disordered analogous to the two-dimensional XY model~\cite{doniach}.
In contrast to microwave measurements of two dimensional films~\cite{pratap},
these measurements~\cite{vlad} suggest phase coherence across the entire 
length of the JJ chain as opposed to puddles. 
 The coherent oscillations in the JJ chain~\cite{vlad} are measured by a 
reflection probe in 
Fig.~\ref{Fig0} that applies an ac electric field with frequency $\omega$ to the end island.
The measured absorption can be related to the imaginary part of the ac polarizability of the
system $\chi(\omega)$ at the end of the JJ chain.
 Here we are ignoring possible power law 
prefactors of $\omega$ that arise from coupling efficiency of $\chi$ to the 
transmission line. Sharp peaks in $\chi(\omega)$ represent resonant 
excitation of a collective mode in the wire. The recent measurement of the 
ac conductivity~\cite{vlad} observes a discrete frequency comb of such peaks 
that suggests excitation of a collective mode associated with phase coherence 
across the insulating chain. 
However, the peaks in the comb appear to 
broaden out and disappear as one goes to lower frequencies, consistent with there being no dc conductance in the insulator.

In this work we calculate the ac polarizability
 $\chi(\omega)$ of an ideal sine-Gordon insulator described by Eq.~\ref{HGiam} near the Luther-Emery 
point (i.e. $K\simeq 1$)~\cite{luther} at vanishingly small temperatures.  
We find that in this limit, the ac polarizability shows sharp oscillations 
associated with the creation of soliton-antisoliton pairs (SAPs). 
Additionally, we use numerical calculations for $K=1$, where the 
model can be mapped to free-fermions, to include disorder 
to show that our conclusions apply qualitatively to the disordered case. Finally, we compare the  
results obtained with the perturbative decay rate of high 
frequency phase modes in the microscopic charge disordered 
JJ chain model. The lifetime of the single-plasmon state is studied by applying second order perturbation theory to quantum phase slips.
This study allows us to put the theoretical results in the context of expectations from  
experimentally realistic superconducting JJ chain that has 
substantial charge disorder as well as a non-linear plasmon dispersion.  

\begin{figure}
\centering
	\includegraphics[width=0.7\textwidth,angle=270]{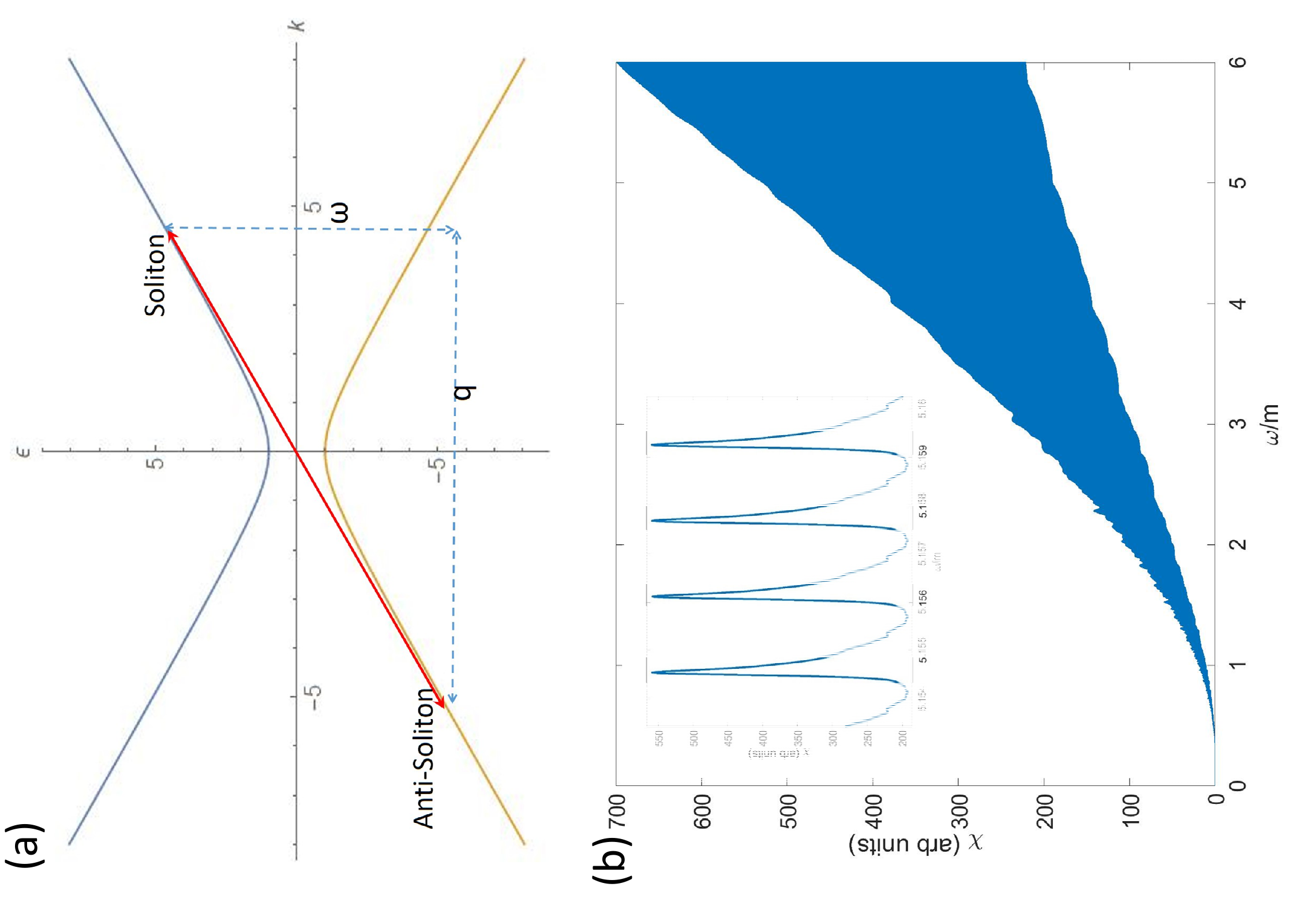}
	\caption{(a) Dispersion ($\varepsilon$ versus $k$) of excitations of the sine-Gordon model with anti-solitons at negative energy. The bold arrow shows charge neutral soliton-antisolition pairs with momentum $q$ and energy $\omega$. Large $q$ 
pairs have a large near-degeneracy with $\omega\simeq q$ constituting a coherent excitation peak.
 (b) Local dynamical polarizability $\chi$ of a sine-Gordon model as a function of $\omega$ shows coherent peaks (see inset) spaced by finite size $1/L$ dominating over 
the incoherent background as frequency increases.
}
	\label{Fig1}
\end{figure}

\section{Soliton-antisoliton pair excitation rates}
The elementary excitations of the sine-Gordon model described by Eq.~\ref{HGiam} are 
solitons (antisolitons) where the phase $\phi$ jumps by $\pm \pi$ between different 
minima of the cosine potential in Eq.~\ref{HGiam}.
Since the phase $\phi$ is 
related to the charge density $\rho(x)\propto\partial_x\phi$, so that 
such solitons (antisolitons) are associated with charge $\pm 2e$. 
A more exact treatment of these apparently static domain walls 
in the sine-Gordon Hamiltonian reveals that such solitons can be viewed as 
essentially non-interacting relativistic quantum particles 
with a dispersion (energy versus momentum relation) that is written as
\begin{align}
&\omega_{s/a}(q)=\pm \sqrt{m^2+q^2 v_c^2}\label{disp}
\end{align} 
and is plotted in Fig.~\ref{Fig1}(a). The index $s/a$ stand for 
solitons and antisolitons respectively. The energy of the anti-soliton is shown 
with a negative sign for convenience of later discussion.  The 
mass parameter is given by 
~\cite{zamolodchikovmass}
\begin{align}
&m=\Upsilon\left[g\frac{\pi\Gamma(1-K/2)}{4\Upsilon\Gamma(K/2)}\right]^{1/(2-K)}\frac{2\Gamma(\xi/2)}{\sqrt{\pi}\Gamma((1+\xi)/2)},\label{mass}
\end{align}
where   $\xi=K/(2-K)$ and $\Upsilon$ is a momentum cut-off scale. 

Let us now discuss qualitatively the origin of the oscillations in the ac polarizability 
$\chi(\omega)$ in terms of the solitons and anti-solitons discussed above.
The polarizability $\chi(\omega)$ is measured by the absorption of photons 
from the probe at the end of the JJ chain (see Fig.~\ref{Fig0}), so that 
$\chi(\omega)$ must be associated with the 
cross-section for generating neutral excitations. 
More formally 
\begin{align}
\chi(\omega)&=\expect{\rho(x;\omega)^\dagger\rho(x;\omega)}=\sum_n |\expect{0|\rho(x)|n}|^2\delta(E_n-\omega)\label{chi0},
\end{align}
where $\rho(x)$ is the charge density operator at the end of the chain.
The state $\ket{0}$ represents the ground state of the system and $\ket{n}$ 
is an excited state with energy $E_n$ above the ground state. 
Such neutral excitations can be constructed by 
pairing the elementary solitons and antisolitons,
which are charged, into  SAPs. 
One such pair with center of mass momentum $q$ is shown in 
Fig.~\ref{Fig1}(a). In this figure, we have flipped the sign of the anti-soliton energy (as mentioned
in the previous paragraph) so that 
 we can extract the energy of the SAP from 
the separation along the axis $\omega_{SAP}(q)=\omega$ (see Fig.~\ref{Fig1}(a) ).
However, the pair shown in Fig.~\ref{Fig1}(a) represents one of a continuum of 
such pairs at momentum $q$, that are parametrized by the relative momentum $k$ so 
that $\ket{n}$ would be the state $\ket{n}\equiv\ket{S_{k+q/2},A_{k-q/2}}$ where $S_{k}$ is the soliton 
with momentum $k$ and $A_k$ is an anti-soliton with momentum $k$.
 Therefore the more general energy of such a pair is given by 
\begin{align}
&\omega_{SAP,k}(q)=\omega_s(k+q/2)-\omega_a(k-q/2).
\end{align}
Clearly the energy of of SAPs $\omega_{SAP,k}(q)$ acquires a natural broadening 
from the dependence on $k$ while $q$ is held fixed.
In fact, the energy of the SAPs at momentum $q$ can take any value $\omega>\sqrt{q^2+4m^2}$ 
by choosing 
\begin{align}
&k=\frac{\omega}{2}\sqrt{\frac{\omega^2-q^2-4m^2}{\omega^2-q^2}}\label{eq:k}.
\end{align}
The density of states (DOS) for the SAPs in the vicinity of a given energy $\omega$, is proportional to $dk/d\omega$, which 
diverges as 
\begin{align}
\frac{dk}{d\omega}\sim (q^2+4m^2)^{4/3}(\omega-\sqrt{q^2+4 m^2})^{-1/2}.\label{dos}
\end{align}
 This leads to a broadened 
peak in optical absorption 
near $\omega\sim\sqrt{q^2+4m^2}$.
This is reminescent of the broadening of the magnon mode in the transverse field Ising model where a magnon fractionalizes into domain walls~\cite{sachdevbook}.
The prefactor $(q^2+4m^2)^{4/3}$ in Eq.~\ref{dos} 
also shows that this peak increases in height as 
$q$ approaches the ultra-relativistic limit $q\gg 2 m$.
The experimental set-up shown in Fig.~\ref{Fig0} 
does not conserve total momentum $q$. In 
contrast, one can excite an array of momenta 
$q=2\pi n/L$, where $L$ is the length of the chain. 
Considering all possible $q$ can expect $\sigma_{ac}(\omega)$ to contain an array of peaks corresponding 
to the divergent DOS of SAPs in Eq.~\ref{dos} 
from each of the allowed values of $q$. In 
principle, one should also consider discrete $k$ 
so that the absorption spectrum should have a 
discrete set of peaks with no broadening. However, 
the spacing of the frequency arising from 
discreteness of $k$ is expected to vanish as one approaches 
the peak of the DOS.

The above qualitative argument ignores the 
matrix elements that would determine the 
absorption cross-sections of the allowed SAPs.
To obtain a more quantitative understanding of the dynamical charge polarizability we compute $\chi(\omega)$ 
directly. Unfortunately, for general values of $K$, the charge operator can couple to multiple SAPs simultaneously (even though we expect this to 
be rare). To avoid this complication we restrict our attention to 
 $K\simeq 1$, which is deep in the insulating phase. At $K=1$,
 the  solitons and antisolitons can be thought of as free Dirac fermions~\cite{coleman,luther,giamarchibook} and 
the charge density operator couples to exactly one soliton-antisoliton pair. This is not necessarily a huge restriction 
because even at $K$ significantly different from $1$, at small values of $g$, the low frequency properties are 
still dominated by $K=1$, which is one of the Luther-Emery fixed points~\cite{giamarchibook,luther}. For the rest 
of this and the next section we set the charge velocity $v_c=1$.
 In the limit $K=1$, the sine-Gordon model Eq.~\ref{HGiam} is equivalent~\cite{coleman} to the 
 massive 1D Dirac model with Hamiltonian
\begin{align}
&H=\sum_k \psi^\dagger_k[k\sigma_z+m\sigma_x]\psi_k,\label{HDirac}
\end{align}
where $\psi_k^\dagger$ are spinors of creation operators for solitons and anti-solitons in Eq.~\ref{HGiam}.
The imaginary part of the local dynamical (i.e. $\omega$ dependent) polarizability  $\chi$ is given by ~\cite{baier1991axial}
\begin{align}
&\chi(\omega)\nonumber\\&\simeq\sum_q \frac{2 m^2 q^2}{(\omega^2-q^2)^{3/2}\sqrt{\omega^2-q^2-4 m^2}}\Theta(\omega^2-q^2-4m^2)\label{chiana}.
\end{align}
The sum over momentum states $q$ takes a discrete set of multiples of $2\pi/L$, where $L$ is the length of the chain. 
This result suggests that the $\chi$ diverges at an array of frequencies near 
$\omega_n^2\simeq (2n\pi/L)^2+4m^2$, where $\chi$ diverges as $(\omega-\sqrt{q^2+4m^2})^{-1/2}$.
In contrast, $\chi$ decays as $\chi\sim 2m^2q^2/\omega^4$ for $\omega\gg q$. Thus, we expect the susceptibility $\chi$ to 
be a broadened set of peaks qualitatively similar to that seen experiment~\cite{vlad}.

This expectation is confirmed from the direct numerical evaluation of Eq.~\ref{chi0} for a finite system as 
 is plotted in Fig.~\ref{Fig1}(b). 
As shown in the inset of Fig.~\ref{Fig1}(b), the shaded areas in 
Fig.~\ref{Fig1}(b) are really a closely spaced set of peaks on an incoherent background. The shape of the peaks 
is consistent with the analytic expectations. 
Considering the shaded areas in Fig.~\ref{Fig1}(b), which elucidates the height of the coherent peaks, we see that 
the height of the oscillations increases with frequency relative to the height of the incoherent background, which is 
consistent with the experiments~\cite{vlad}.
For the numerical evaluation, we have broadened the delta functions in Eq.~\ref{chi0} by $0.1(2\pi/L)$ (where $L$ is the chain length), which is much smaller than the plasmon level spacing. The value of the broadening does not appear to affect the result as long as it is smaller than the peak spacing since 
$o(L)$ SAPs contribute to each peak that are spaced by $o(1/L)$. 
Thus we do not expect the broadening to affect the shape of the peaks as long as it is larger than $o(1/L^2)$.

\begin{figure}
\centering
	\includegraphics[width=0.5\textwidth,angle=0]{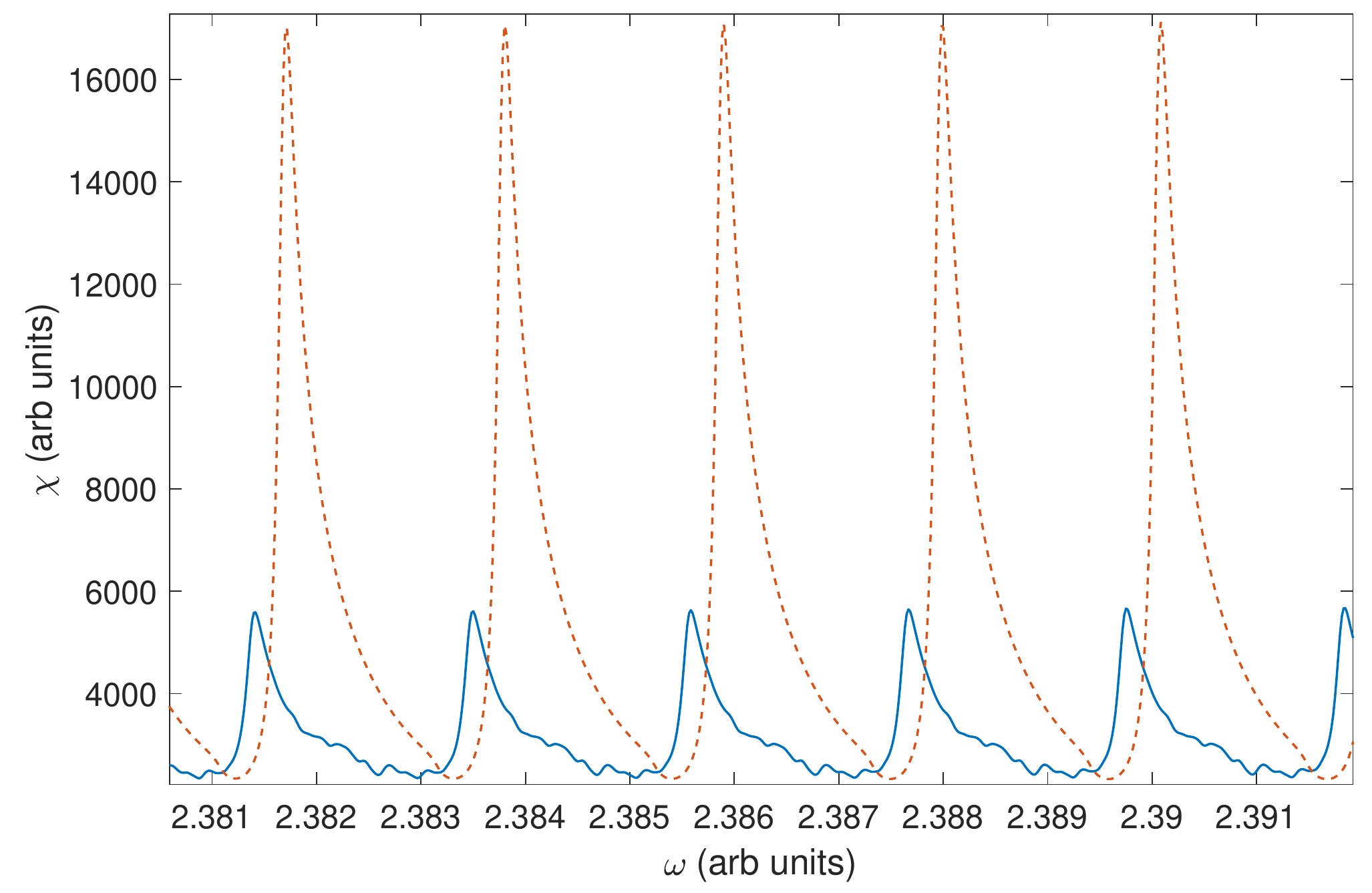}
	\caption{(Color online) Luttinger parameter $K$ dependence of resistance oscillations. 
Change in Luttinger parameter decreases the mass gap $m$ and enhances the 
oscillation for $K=1.4$ (red dashed line). The solid blue curve shows suppressed 
oscillations at $K=1$ since the frequency is of order mass. 
}
	\label{Fig1.5}
\end{figure}
  
The discussion above for $K=1$ also applies for $K\gtrsim 1$. In this case the excitation process in the non-interacting 
massive Dirac model is replaced by the form factor for creating an unbound SAPs~\cite{mandelstam}. 
For the general interacting case, 
contributions from multiple pairs must be considered. However, such contributions are likely  to be small at low energies because 
of phase space constraints as well as weak interactions near $K\sim 1$. 
The dominant effect of changing the Luttinger paramter $K$ is to renormalize the mass downwards with 
increasing $K$ following Eq.~\ref{mass}. Reducing the mass $m$, effectively increases the dimensionless frequency $\omega/m$,
which determines the height of peaks in the dynamical polarizability $\chi$. 
This suggests, following the results in Fig.~\ref{Fig1}(b), that increasing the Luttinger parameter $K$ enhances the strength 
of oscillations in $\chi$ for a fixed frequency. 
 
An additional contribution to the $K$ dependence of the dynamical charge polarizability $\chi$ is the 
  matrix element for the charge density operator $\rho$. Specifically, Bethe ansatz methods have been used to show 
that this matrix element is modified by a factor~\cite{babujian}
\begin{align}
f_-(\vartheta)=&\frac{\cosh\frac{i\pi-\vartheta}{2}}{2\cosh\frac{(i\pi-\vartheta)K}{2}} \nonumber\\
&\times \exp\int^\infty_0 \frac{dt}{t}\frac{\sinh\frac{(K-1)t}{2K}[1-\cosh t(1+\frac{i\vartheta}{\pi})]}{\sinh (t)\sinh(\frac{t}{2K})\cosh(\frac{t}{2})}\label{eqf}
\end{align} 
and $\vartheta_j$ are the rapidities of the solitons that are defined by the equation 
$k_j=m\sinh{\vartheta_j}$.
For weakly attractive fermions, $K\gtrsim 1$, the interaction part of the form factor $f\sim \vartheta$ for small $\vartheta$ and $f\sim e^{\alpha\vartheta}$
for $\vartheta\gg 1$,
where $\alpha$ is a $K$ dependent constant. For $q\gg m$, $\vartheta\sim \log{q}$ so $f\propto q^{2\alpha}$. This leads to 
a $K$ dependent power law for the increasing of the absorption peaks as one goes to more strongly attractive fermions i.e. towards the superfluid phase.

The dynamical polarizability $\chi$ at $K\gtrsim 1$, computed using Eq.~\ref{chi0} 
within the one SAP approximation described above, 
 is plotted in Fig.~\ref{Fig1.5}. Consistent with the theoretical expectation, Fig.~\ref{Fig1.5} 
shows the suppression of oscillations as one  increases the Luttinger parameter $K$ towards the 
the superfludi phase. 
  The results for deep in the insulating phase (i.e. $K\sim 1$),
which are shown in the solid blue curve, shows suppressed oscillations 
consistent with recent experiments~\cite{vlad}. In 
comparison the red dashed curve, which is less insulating (i.e. $K\sim 1.4$), shows 
much stronger oscillations. As discussed earlier, the one SAP approximation  
 cannot be used to consider Luttinger parameter $K$ far from the 
non-interacting point $K=1$.

\section{Disordered massive Dirac model} 
Let us now consider the effect of weak potential disorder, 
which is an intrinsic part of the JJ chain set-up. As will be elaborated in the next section,
the long range nature of the Coulomb interaction converts the uncorrelated random background
charge into a smooth  potential disorder. This allows the disorder potential to be consistent with the low-energy and long-wavelength limit required for the 
applicability of the sine-Gordon model.  
 In this section, we restrict our analysis to $K\sim 1$  
where we can map the sine-Gordon model with disorder potential to the 
massive Dirac model~\cite{coleman}:
\begin{align}
&H=\int dx \psi^\dagger(x)[i \partial_x\sigma_z+m\sigma_x-\mu(x)]\psi(x),\label{Hthirring}
\end{align}
where $\psi^\dagger(x)$ are the Fourier transform of the spinors in Eq.~\ref{HDirac}. The disorder potential is included 
in $\mu(x)$, which is an uncorrelated potential that is assumed to be smooth on
 the scale of the 
spacing of the JJ chain. 
The fluctuations in the local potential $\mu(x)$ leads to back-scattering of fermions 
at energy $E$, which leads to a mean free path (Appendix C): 
\begin{align}
\lambda = \frac{4E^2+V_0^4\Lambda^2}{2m^2V_0^2\Lambda},\label{lambda}
\end{align}
where $V_0^2\Lambda$ characterizes the strength of the fluctuations in $\mu(x)$ at length scale according 
to the relation $\langle \mu(x)\mu(x')\rangle = V_0^2\Lambda\delta(x-x')$.
The mean-free path increases from $V_0^2\Lambda/2 m^2$ to $2E^2/ m^2 V_0^2 \Lambda$ as the 
energy $E$ of the excited soliton/antisoliton increases from $0$. Such a chain appears insulating for lengths 
that are longer than $L\gg V_0^2\Lambda/2 m^2$. Note that the effective length diverges if the mass $m$ 
drops to zero as is expected to occur near the superfluid phase from Eq.~\ref{mass}. Furthermore,
  we expect SAPs to be localized at low frequency and not contribute to resonant 
excitations. 
In contrast, the dynamical polarizability $\chi$ at frequencies above $\omega\gg m V_0\sqrt{\Lambda L/2}$, 
is dominated by high energy delocalized (see Eq.~\ref{lambda}) SAPs and  should 
show sharp resonances similar to the clean result in Fig.~\ref{Fig1}.
\begin{figure}
\centering
	\includegraphics[width=0.5\textwidth,angle=0]{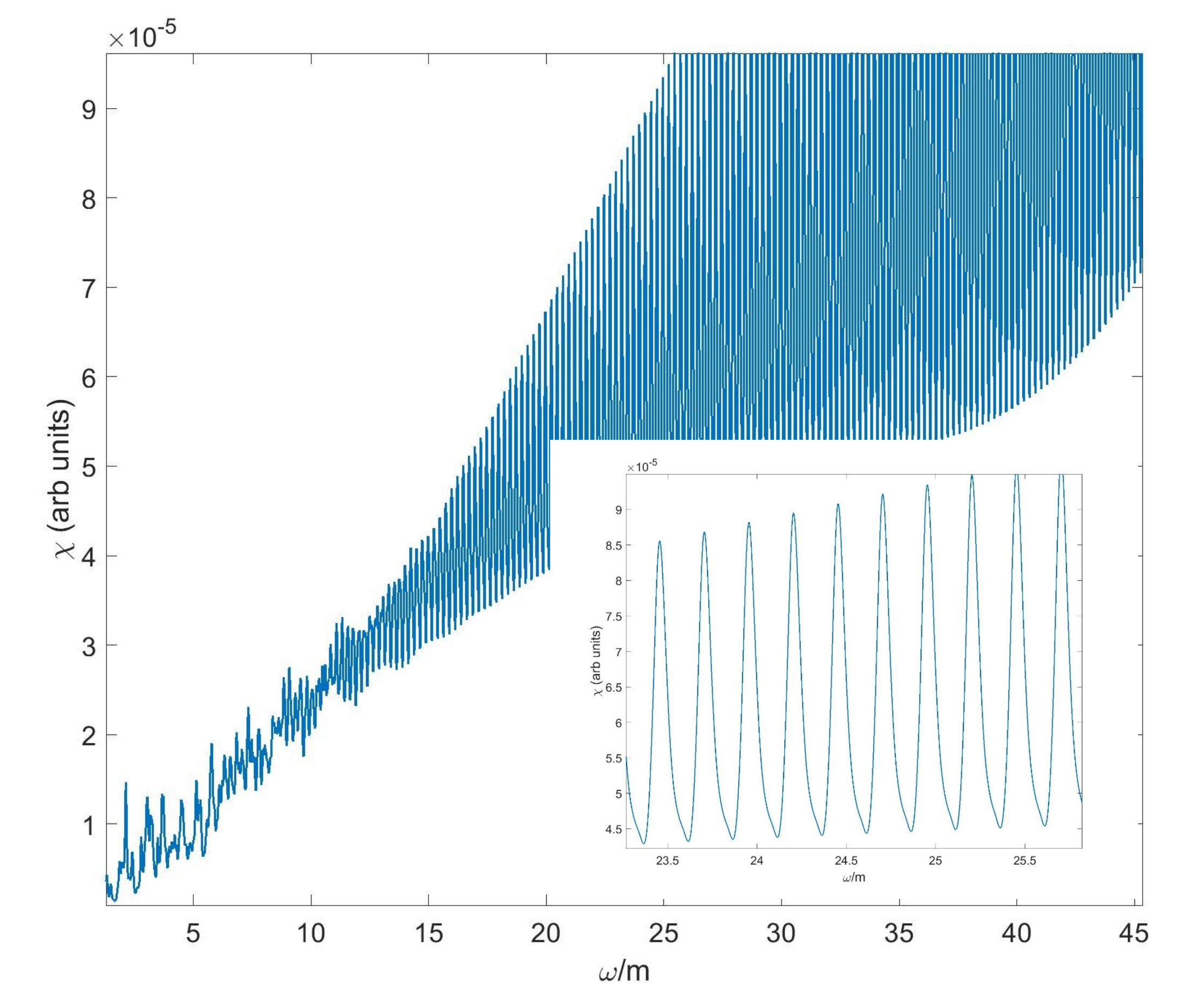}
	\caption{Local dynamical polarizability $\chi$ of the disordered massive Dirac model as a function of $\omega$ shows coherent peaks at high frequencies and smaller random absorption peaks at low frequencies. Inset shows that the peak shape in the disordered 
system, which is more symmetric compared to that in the clean system shown in the inset in Fig.~\ref{Fig1.5}.
}
	\label{Fig2}
\end{figure}

These expectations can be verified from $\chi$, which is calculated by a numerical evaluation of Eq.~\ref{chi0} for the Hamiltonian 
Eq.~\ref{Hthirring} and is  
 plotted in Fig.~\ref{Fig2}. The result shows a disorder gap at low frequency and plasma oscillations at high frequency. 
Consistent with the intuitive argument at the beginning of this section, high frequency SAPs that 
contribute to $\chi$ with frequencies $\omega\gg m$ are not scattered by the disorder and therefore lead to coherent oscillations in 
$\chi(\omega)$ seen in Fig.~\ref{Fig2}. On the other hand, the low energy SAPs are pinned by the disorder domain walls. 
The low energy excitation constitutes bound resonances from exciting vibration modes of such bound solitons. Such 
solitons being localized and random appear as a sequence of random peaks in Fig.~\ref{Fig2} that are qualitatively different 
from Fig.~\ref{Fig1}(b). 

Despite the fact that such disorder breaks integrability, we can generalize these arguments to $K\gtrsim 1$. For a 
general sine-Gordon model (Eq.~\ref{HGiam}),  weak potential disorder works as introducing a pinning 
potential for the massive solitons. For $K<3/2$ and weak disorder, 
such potentials lead to pinning of all low energy solitons leading to an insulating phase~\cite{giamarchischulz}. 
Higher energy solitons, which can be obtained by applying a Lorentz boost, have shorter length scales and are therefore not 
scattered by smooth charge disorder. The coherent high energy SAPs can  generate large coherent oscillations in the 
dynamical polarizability $\chi$ at high frequencies. 

\section{Nearly superfluid limit of the JJ chain}\label{nearSF}
We now consider the case where the array is closer to the superfluid phase. In this limit, the 
Luther-Emery model (Eq.~\ref{HDirac}) is no longer valid. 
Furthermore, the sine-Gordon model Eq.~\ref{HGiam} ignores curvature effects in the energy.
 However, the effect of phase slips (i.e. $g$ in Eq.~\ref{HGiam} ) 
is perturbatively weak in this limit - so that one can compute the decay directly from the microscopic model 
for a JJ chain that includes both island capacitance $C_g$, JJ capacitance $C_J$ as well as charge disorder. 
In this section, we consider a more microscopically justified Hamiltonian for the system that is written as:
\begin{align}
H = &\sum\limits_{i,j} U_{i,j}(N_i-Q_i)(N_j-Q_j) \nonumber\\&+ E_J\sum\limits_i[1-\cos(\theta_{i}-\theta_{i-1})] \label{eqH}
\end{align}
where $N_i$ and $\theta_i$ are number of Cooper pairs and phase at site i, $E_J$ is the Josephson energy of junctions and
\begin{align}
U_{i,j} = \frac{1}{M}\sum\limits_q U_q e^{iq(R_i-R_j)}
\end{align}
where $M$ is the system size and $U_q = 4E_0E_1/[E_1 + 4E_0\sin^2(q/2)]$, charging energies $E_{0,1}$ are defined as
 $E_0 = e^2/2C_g$ and $E_1 = e^2/2C_J$. 
$\{Q's\}$ are normally distributed random stray charges which satisfy $\langle \langle Q_iQ_j \rangle\rangle = D\delta_{i,j}$.
The corresponding random potential $\mu_i=\sum_j U_{i,j}Q_j$, has a Fourier transform $$\langle\langle |\mu_q|^2\rangle\rangle \approx D\frac{16 E_0^2 E_1^2}{(E_1+E_0 q^2)^2},$$
 which is strongly peaked at $k\sim 0$ for $E_0\gg E_1$. 
Thus the potential in this case can be assumed to be smooth as assumed in Eq.~\ref{Hthirring}.

For energy of states below the Josephson energy $E_J$, we expect  $(\theta_i-\theta_{i-1})\ll 2\pi$. However, 
at low energies rare events called phase-slips locally shift one of the phase differences $(\theta_i-\theta_{i-1})$ by 
$2\pi$. Such a phase slip is local i.e. it doesn't affect the phase $\theta_j$ for $|j-i|\gg 1$. This implies that the 
remaining of the phase differences add up to $2\pi$ immediately following the phase slip.
The phase-slip operator is simpler to represent in dual variables defined as 
\begin{align}
&\Pi_j=(\theta_{j+1}-\theta_j)/\pi\\
&\phi_j=\pi\sum_{l\leq j}N_l.
\end{align}
The commutation relation of these operators 
\begin{align}
&[\phi_l, \Pi_j]=\sum_{m\leq l}[N_m, \theta_{j+1}-\theta_j]\\
&=i\sum_{m\leq l}(\delta_{j-m}-\delta_{j-m+1})=i\delta_{j-l}.
\end{align}
A quantum phase slip at site $j$ is created by operators $e^{\pm 2i\bar{\phi}_j}$, where 
\begin{align}
&\bar{\phi}_j=\sum_l w_{l-j}\phi_l,
\end{align}
where $w_l\geq 0$ are normalized weights which peak at $l=0$ so that $\sum_l w_l=1$. The width of $w_l$ 
represents the length scale of the phase slip (Appendix B). Thus, phase-slips can be considered to be nucleated by a 
term $g\sum_j\cos{2\bar{\phi}_j}$. We can write the
the low-energy effective Hamiltonian where the phase difference variables $\Pi_j\ll 1$, including the phase-slip 
term as 
\begin{align}
H = &\sum\limits_{i,j} \frac{U_{i,j}}{\pi^2}(\phi_{i}-\phi_{i-1}-\pi Q_i)(\phi_{j}-\phi_{j-1}-\pi Q_j) \nonumber\\&+ \frac{\pi^2E_J}{2}\sum\limits_i\Pi_{i}^2+g\sum_i \cos{2\bar{\phi}_i}. \label{eqHSG}
\end{align}
The disorder from $Q_j$ in the charging energy can be eliminated by a unitary transformation $e^{-i\pi\sum_j Q_j \Pi_j}$ which shifts $\bar{\phi}_j$ to $\bar{\phi}_j+ \Lambda_j$, where $\Lambda_j = \pi\sum_l w_{l-j}\sum_{m = 1}^l Q_m$.

Ignoring $g$ for the moment, diagonalizing $H$ under open boundary condition $\Pi_1 = \Pi_{M-1} = 0$ gives rise to sound-like plasmonic excitation ($\hbar = 1$),
\begin{align}
H_0 &= \sum_q \frac{\pi^2E_J}{2}\Pi_q^2+ \frac{4\sin^2(q/2)U_q}{\pi^2}\phi_q^2 \nonumber\\
&= \sum\limits_{q} \Omega_q (a^\dagger_qa_q+\frac{1}{2}),
\end{align}
where $q$ takes integer multiples of $\pi/M$, and
\begin{align}
\Omega_q = \sqrt{\frac{32E_0E_1E_J}{E_1+4E_0\sin^2(q/2)}}\sin(q/2)
\end{align}
is the dispersion of the plasmonic mode that becomes linear ($\sim v_cq$) with speed $v_c = \sqrt{8E_0E_J}$ as $q$ goes to $0$ and 
reaches a maximum at the plasma frequency $\omega_p = \sqrt{8E_1E_J}$. 

\begin{figure}
\centering
	\includegraphics[width=0.45\textwidth]{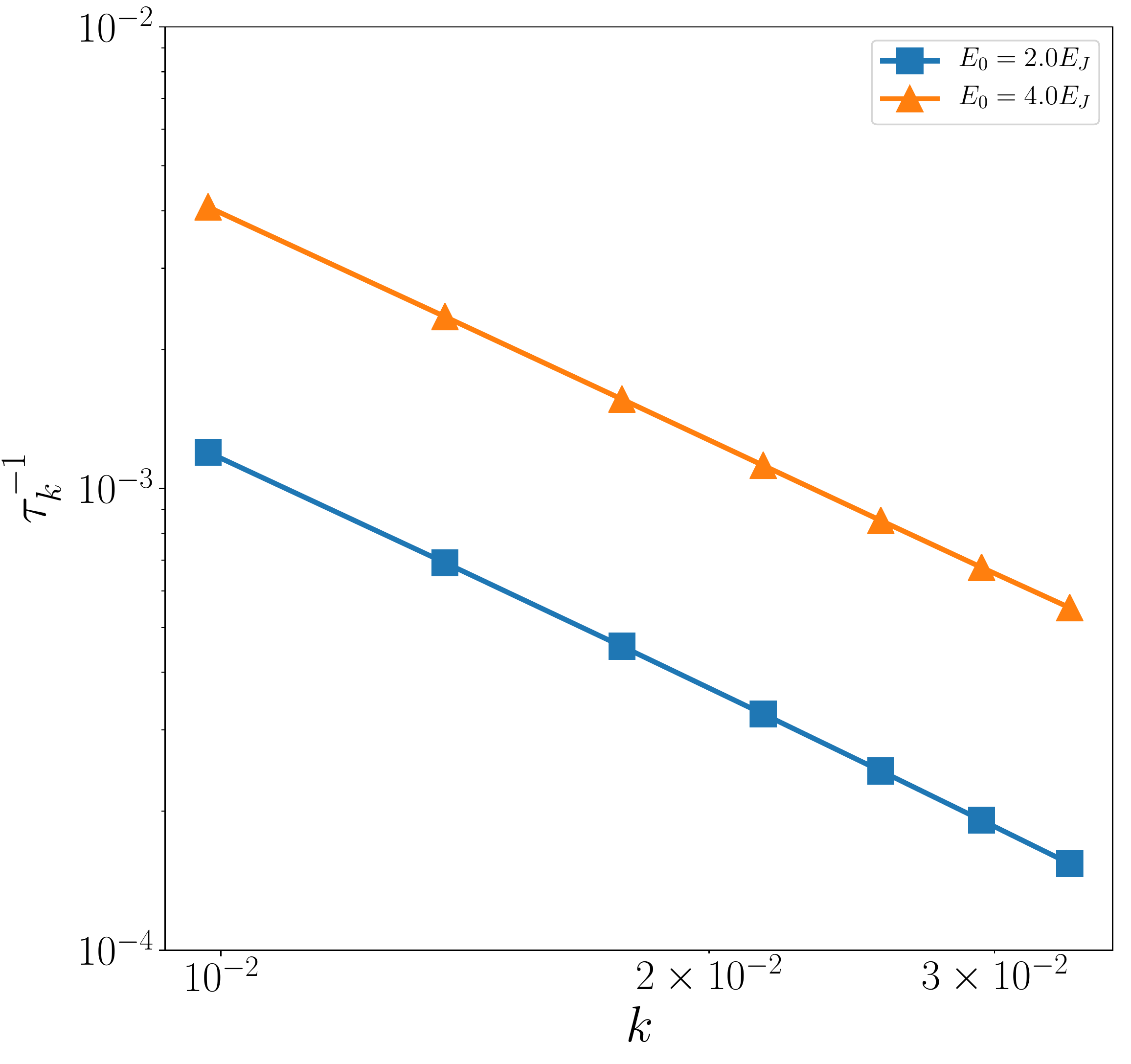}
	\caption{Inverse lifetime of single plasmon states scaled by $g^2/E_J$. The system parameters are $M=1600, D = 0.02,E_1/E_J = 0.1$ while $E_0$ is changed. Power-law increase of the decay rate at smaller wave vectors ( i.e. $q \rightarrow 0$) is consistent with weaker peaks at smaller frequencies obtained from the sine-Gordon model. Moreover, the decay rate is suppressed as $E_0$ decreases, which is consistent with the experiment \cite{vlad}.}
	\label{Fig:DecayRate}
\end{figure}

At finite $g\neq 0$, phase-slips couple single-plasmon state $\vert k \rangle \equiv a_k^\dagger \vert 0\rangle$ to multi-plasmon states. To get the plasmon lifetime, we will consider its self-energy, which, to lowest order, is described by the cubic term in the normal-ordered interaction
\begin{align}
\frac{1}{3!}g\sum_i \xi_i \sin2\Lambda_i:(2\bar{\phi}_i)^3:.
\end{align}
The additional factor
\begin{align}
\xi_i = \exp[-\frac{\pi^2}{8}\sum_q\sqrt{\frac{E_J}{2\sin ^2(q/2)U_q}} (A^i_q)^2] 
\end{align}
arises from normal-ordering defined by $\cos2\bar{\phi}_i = \xi_i :\cos2\bar{\phi}_i:$ where $A^i_q = \sqrt{8/M}\sum_l w_{l-i}\sin ql$ is defined to be the coefficient in $2\bar{\phi}_i = \sum_q A^{i}_q\phi_q$.

The lifetime of momentum $k$ plasmon obtained by self-energy calculation to order $g^2$ is given by (see Appendix D for details).
\begin{align}
\frac{1}{\tau_k} = \pi g^2\sum_{q_1,q_2}(\Gamma_{k,q_q,q_2})^2\frac{\delta(\Omega_k-\Omega_{q_1}-\Omega_{q_2})}{8\Omega_k\Omega_{q_1}\Omega_{q_2}}
\end{align}
where the matrix element is 
\begin{align}
\Gamma_{q_1,q_2,q_3} =  (\pi\sqrt{E_J})^{3}\sum_i\xi_i\sin 2\Lambda_i A^i_{q_1}A^i_{q_2}A^i_{q_3}.
\end{align}

Next, we perform disorder average on the inverse lifetime, 
\begin{align}
\langle\langle \frac{1}{\tau_k}\rangle\rangle = \pi g^2\sum_{q_1,q_2}\langle\langle(\Gamma_{k,q_q,q_2})^2\rangle\rangle\frac{\delta(\Omega_k-\Omega_{q_1}-\Omega_{q_2})}{8\Omega_k\Omega_{q_1}\Omega_{q_2}}.
\label{eq:avg_damping}
\end{align}
Using the fact that the correlation function of the Aharonov-Casher phase shift, $\langle\langle e^{2i(\Lambda_i-\Lambda_j)}\rangle\rangle = e^{-2\pi^2D\sum_l(w_{l-i}-w_{l-j})^2}$,
\begin{align}
\langle\langle(\Gamma_{k,q_q,q_2})^2\rangle\rangle =& \frac{(\pi^2E_J)^3}{2}\sum_{i,j}\xi_i\xi_j e^{-2\pi^2D\sum_l(w_{l-i}-w_{l-j})^2} \nonumber\\&\times(A^i_{k}A^i_{q_1}A^i_{q_2})(A^j_{k}A^j_{q_1}A^j_{q_2})
\end{align}

The inverse lifetime for single-plasmon states evaluated using Eq.~\ref{eq:avg_damping} is shown in Fig.~\ref{Fig:DecayRate}. The phase slip profile is chosen such that $w_{l} = \{\tanh[0.2(l-1/2)] - \tanh[0.2(l-3/2)]\}/2$, which expands through a length scale of 5 sites. System parameters are chosen to be $M = 1600$, $D = 0.02$ and $E_1/E_J = 0.1$ while $E_0/E_J$ is tuned. Next, we fix the factor $\xi_i$'s, which are expected to be a constant of $i$ in the $M\rightarrow \infty$ limit, to be $\xi_{M/2}$. This would avoid some finite size effects. Finally, to obtain a finite result from Eq.~\ref{eq:avg_damping}, we introduce a Gaussian broadening $\eta = 0.5\omega_{q_{min}}$ in the energy conservation delta function. Such a broadening is essential to obtain a result from the zero temperature approach for a finite system with curvature in the dispersion relation. This issue may be 
circumvented at finite temperature, where the broadening is created self-consistently using the decay rate. However, such a treatment goes beyond systematic the 
systematic perturbation theory in the phase-slip amplitude $g$ considered here.

\section{Conclusions and discussion}
In this work we have calculated the imaginary part of the local dynamical 
polarizability $\chi$ in the limit where the Luttinger parameter $K\sim 1$ (corresponding to 
impedance of order of a conductance quantum) where $\chi$ is dominated by the cross-section 
of generation of SAPs. We find that the resulting $\chi$, for $K=1$, shows an array of peaks (see Eq.~\ref{chiana})
whose height increases with increasing frequency in a way that is qualitatively consistent with recent experimental 
measurements~\cite{vlad}. Increasing 
$K$ towards the superfluid limit reduces the effective mass (see Eq.~\ref{mass}) of the solitons leading to larger oscillations.
Such oscillations in $\chi$ are surprising at first glance because a JJ chain 
in the insulating phase does not have superconducting phase coherence.
We show that the observed oscillations are essentially a consequence of the Lorentz invariance of 
the sine-Gordon model description of the chain (Eq.~\ref{HGiam}).  The coherent phase oscillations in this model 
are gapped by a Lorentz-invariant  phase-slip (cosine) term,  for weak Josephson 
coupling $E_J$, resulting in an insulating JJ chain.
 Such a sine-Gordon model turns out to be a
good approximation for long JJ chains in the limit of large JJ capacitance~\cite{choi1998quantum}.
This is because the Luttinger parameter $K$ (inverse impedance) and the phase-slip amplitude $g$ are independently controlled by 
the long and short ranged part of the effective screened Coulomb interaction in the micoscopic model (see Sec.~\ref{nearSF}). 
Such clean sine-Gordon models have been proposed to be realized in 
strongly interacting ultra-cold Bose gases~\cite{zwerger}. 
We expect similar broadened above gap oscillations in $\chi$ to apply to a superfluid-insulator transition in the context of ultra-cold atoms~\cite{zwerger}.

 The coherent phase peaks in $\chi$ are also found to be 
robust to charge disorder that is present in JJ chains~\cite{vlad}. The long-range 
nature of the Coulomb interactions that are encoded in the capacitances in the 
JJ chain model ensures that the uncorrelated charge disorder appears smooth 
on the lattice scale, ensuring the validity of the sine-Gordon model treatment.
 We have shown analytically that the mean-free path for solitons increases with frequency for SAPs. 
However, as the frequency of the perturbation is lowered 
the mean-free path of the SAPs become short leading to localized excitations.
 This leads to increased broadening of the resonances at lower frequency consistent with 
recent experiments~\cite{vlad}. We also find that the resonances rest on 
an incoherent background also seen in the experimental data 
for high impedance wires~\cite{vlad}. Additionally, the models considered here all 
have a gap leading to insulating transport in dc, which is also consistent with 
the recent experiments~\cite{vlad}. The insulating behavior may be 
understood as charge disorder pinning of the 
charged solitons~\cite{giamarchischulz}. 

 In the last section of this work, we consider the perturbative 
decay of plasmons due to quantum phase slips, similar to recent work~\cite{mirlindecay}. However, in contrast to previous work~\cite{mirlindecay},
we avoid analytic continuation and directly 
 evaluate the self-energy with the quantum phase slip term in Eq.~\ref{HGiam}
being treated as a perturbation. The analytic continuation is particularly difficult for finite size and disordered system.
 Furthermore, we explicitly exclude disorder broadening 
of single-plasmon states. Such broadening is implicitly included in the replica-based 
disorder averaged partition function approach~\cite{mirlindecay}.
However, as shown in recent work~\cite{HouzetGlazman}, disorder-induced fluctuations can also lead to extrinsic broadening 
of the resonances. 
We also find that strictly speaking, such a perturbative decay maybe 
limited or artificially enhanced by energy conservation in a finite chain. 
We formally circumvent this problem by introducing an intrinsic broadening of the states.
We find that the calculated decay rates that increase as one reduces 
wave-vector $q$ consistent with both experiment~\cite{vlad} and 
recent theory~\cite{mirlindecay}. However, it is not clear that the intrinsic source of broadening is justified at the 
low temperatures required to have an insulator.

Finally, we note that our work relies on the integrability of the sine-Gordon 
model in the clean limit. We have also assumed that this integrability extends 
to the disordered sine-Gordon model near the Luther-Emery 
point~\cite{luther,giamarchibook} ($K\simeq 1$) to describe finite real frequency 
dynamics. On the other hand, the general disordered sine-Gordon model is 
not integrable and likely exhibits interesting dynamical behavior related to 
many-body localization~\cite{mblreview}. This likely leads to avenues for 
interesting work on this system in the future regarding low frequency equilibration 
at this system.

We thank Vladimir Manucharyan for introducing us to this problem and 
M. Houzet for valuable discussions.
This work was supported by the  NSF-DMR-1555135 (CAREER), JQI-NSF-PFC (PHY1430094)
and the Sloan research fellowship.

\bibliographystyle{apsrev4-1}
\bibliography{ThirringModelabsorption0517}

\begin{thebibliography}{30}%
\makeatletter
\providecommand \@ifxundefined [1]{%
 \@ifx{#1\undefined}
}%
\providecommand \@ifnum [1]{%
 \ifnum #1\expandafter \@firstoftwo
 \else \expandafter \@secondoftwo
 \fi
}%
\providecommand \@ifx [1]{%
 \ifx #1\expandafter \@firstoftwo
 \else \expandafter \@secondoftwo
 \fi
}%
\providecommand \natexlab [1]{#1}%
\providecommand \enquote  [1]{``#1''}%
\providecommand \bibnamefont  [1]{#1}%
\providecommand \bibfnamefont [1]{#1}%
\providecommand \citenamefont [1]{#1}%
\providecommand \href@noop [0]{\@secondoftwo}%
\providecommand \href [0]{\begingroup \@sanitize@url \@href}%
\providecommand \@href[1]{\@@startlink{#1}\@@href}%
\providecommand \@@href[1]{\endgroup#1\@@endlink}%
\providecommand \@sanitize@url [0]{\catcode `\\12\catcode `\$12\catcode
  `\&12\catcode `\#12\catcode `\^12\catcode `\_12\catcode `\%12\relax}%
\providecommand \@@startlink[1]{}%
\providecommand \@@endlink[0]{}%
\providecommand \url  [0]{\begingroup\@sanitize@url \@url }%
\providecommand \@url [1]{\endgroup\@href {#1}{\urlprefix }}%
\providecommand \urlprefix  [0]{URL }%
\providecommand \Eprint [0]{\href }%
\providecommand \doibase [0]{http://dx.doi.org/}%
\providecommand \selectlanguage [0]{\@gobble}%
\providecommand \bibinfo  [0]{\@secondoftwo}%
\providecommand \bibfield  [0]{\@secondoftwo}%
\providecommand \translation [1]{[#1]}%
\providecommand \BibitemOpen [0]{}%
\providecommand \bibitemStop [0]{}%
\providecommand \bibitemNoStop [0]{.\EOS\space}%
\providecommand \EOS [0]{\spacefactor3000\relax}%
\providecommand \BibitemShut  [1]{\csname bibitem#1\endcsname}%
\let\auto@bib@innerbib\@empty
\bibitem [{\citenamefont {Nandkishore}\ and\ \citenamefont
  {Huse}(2015)}]{mblreview}%
  \BibitemOpen
  \bibfield  {author} {\bibinfo {author} {\bibfnamefont {R.}~\bibnamefont
  {Nandkishore}}\ and\ \bibinfo {author} {\bibfnamefont {D.~A.}\ \bibnamefont
  {Huse}},\ }\href {\doibase 10.1146/annurev-conmatphys-031214-014726}
  {\bibfield  {journal} {\bibinfo  {journal} {Annual Review of Condensed Matter
  Physics}\ }\textbf {\bibinfo {volume} {6}},\ \bibinfo {pages} {15} (\bibinfo
  {year} {2015})}\BibitemShut {NoStop}%
\bibitem [{\citenamefont {Sachdev}(2012)}]{adscft}%
  \BibitemOpen
  \bibfield  {author} {\bibinfo {author} {\bibfnamefont {S.}~\bibnamefont
  {Sachdev}},\ }\href@noop {} {\bibfield  {journal} {\bibinfo  {journal} {Annu.
  Rev. Condens. Matter Phys.}\ }\textbf {\bibinfo {volume} {3}},\ \bibinfo
  {pages} {9} (\bibinfo {year} {2012})}\BibitemShut {NoStop}%
\bibitem [{\citenamefont {Sachdev}(2011)}]{sachdevbook}%
  \BibitemOpen
  \bibfield  {author} {\bibinfo {author} {\bibfnamefont {S.}~\bibnamefont
  {Sachdev}},\ }\href@noop {} {\emph {\bibinfo {title} {Quantum phase
  transitions}}}\ (\bibinfo  {publisher} {Cambridge university press},\
  \bibinfo {year} {2011})\BibitemShut {NoStop}%
\bibitem [{\citenamefont {Fisher}(1990)}]{fisher}%
  \BibitemOpen
  \bibfield  {author} {\bibinfo {author} {\bibfnamefont {M.~P.~A.}\
  \bibnamefont {Fisher}},\ }\href {\doibase 10.1103/PhysRevLett.65.923}
  {\bibfield  {journal} {\bibinfo  {journal} {Phys. Rev. Lett.}\ }\textbf
  {\bibinfo {volume} {65}},\ \bibinfo {pages} {923} (\bibinfo {year}
  {1990})}\BibitemShut {NoStop}%
\bibitem [{\citenamefont {Markovi{\'c}}\ \emph {et~al.}(1998)\citenamefont
  {Markovi{\'c}}, \citenamefont {Christiansen},\ and\ \citenamefont
  {Goldman}}]{goldman}%
  \BibitemOpen
  \bibfield  {author} {\bibinfo {author} {\bibfnamefont {N.}~\bibnamefont
  {Markovi{\'c}}}, \bibinfo {author} {\bibfnamefont {C.}~\bibnamefont
  {Christiansen}}, \ and\ \bibinfo {author} {\bibfnamefont {A.}~\bibnamefont
  {Goldman}},\ }\href@noop {} {\bibfield  {journal} {\bibinfo  {journal} {Phys.
  Rev. Lett.}\ }\textbf {\bibinfo {volume} {81}},\ \bibinfo {pages} {5217}
  (\bibinfo {year} {1998})}\BibitemShut {NoStop}%
\bibitem [{\citenamefont {Mason}\ and\ \citenamefont
  {Kapitulnik}(2002)}]{kapitulnik}%
  \BibitemOpen
  \bibfield  {author} {\bibinfo {author} {\bibfnamefont {N.}~\bibnamefont
  {Mason}}\ and\ \bibinfo {author} {\bibfnamefont {A.}~\bibnamefont
  {Kapitulnik}},\ }\href@noop {} {\bibfield  {journal} {\bibinfo  {journal}
  {Phys. Rev. B}\ }\textbf {\bibinfo {volume} {65}},\ \bibinfo {pages} {220505}
  (\bibinfo {year} {2002})}\BibitemShut {NoStop}%
\bibitem [{\citenamefont {Hebard}\ and\ \citenamefont
  {Paalanen}(1990)}]{hebard}%
  \BibitemOpen
  \bibfield  {author} {\bibinfo {author} {\bibfnamefont {A.}~\bibnamefont
  {Hebard}}\ and\ \bibinfo {author} {\bibfnamefont {M.}~\bibnamefont
  {Paalanen}},\ }\href@noop {} {\bibfield  {journal} {\bibinfo  {journal}
  {Phys. Rev. Lett.}\ }\textbf {\bibinfo {volume} {65}},\ \bibinfo {pages}
  {927} (\bibinfo {year} {1990})}\BibitemShut {NoStop}%
\bibitem [{\citenamefont {Mondal}\ \emph {et~al.}(2013)\citenamefont {Mondal},
  \citenamefont {Kamlapure}, \citenamefont {Ganguli}, \citenamefont
  {Jesudasan}, \citenamefont {Bagwe}, \citenamefont {Benfatto},\ and\
  \citenamefont {Raychaudhuri}}]{pratap}%
  \BibitemOpen
  \bibfield  {author} {\bibinfo {author} {\bibfnamefont {M.}~\bibnamefont
  {Mondal}}, \bibinfo {author} {\bibfnamefont {A.}~\bibnamefont {Kamlapure}},
  \bibinfo {author} {\bibfnamefont {S.~C.}\ \bibnamefont {Ganguli}}, \bibinfo
  {author} {\bibfnamefont {J.}~\bibnamefont {Jesudasan}}, \bibinfo {author}
  {\bibfnamefont {V.}~\bibnamefont {Bagwe}}, \bibinfo {author} {\bibfnamefont
  {L.}~\bibnamefont {Benfatto}}, \ and\ \bibinfo {author} {\bibfnamefont
  {P.}~\bibnamefont {Raychaudhuri}},\ }\href@noop {} {\bibfield  {journal}
  {\bibinfo  {journal} {Scientific reports}\ }\textbf {\bibinfo {volume} {3}},\
  \bibinfo {pages} {1357} (\bibinfo {year} {2013})}\BibitemShut {NoStop}%
\bibitem [{\citenamefont {Greiner}\ \emph {et~al.}(2002)\citenamefont
  {Greiner}, \citenamefont {Mandel}, \citenamefont {Esslinger}, \citenamefont
  {H{\"a}nsch},\ and\ \citenamefont {Bloch}}]{greiner}%
  \BibitemOpen
  \bibfield  {author} {\bibinfo {author} {\bibfnamefont {M.}~\bibnamefont
  {Greiner}}, \bibinfo {author} {\bibfnamefont {O.}~\bibnamefont {Mandel}},
  \bibinfo {author} {\bibfnamefont {T.}~\bibnamefont {Esslinger}}, \bibinfo
  {author} {\bibfnamefont {T.~W.}\ \bibnamefont {H{\"a}nsch}}, \ and\ \bibinfo
  {author} {\bibfnamefont {I.}~\bibnamefont {Bloch}},\ }\href@noop {}
  {\bibfield  {journal} {\bibinfo  {journal} {nature}\ }\textbf {\bibinfo
  {volume} {415}},\ \bibinfo {pages} {39} (\bibinfo {year} {2002})}\BibitemShut
  {NoStop}%
\bibitem [{\citenamefont {Podolsky}\ and\ \citenamefont
  {Sachdev}(2012)}]{podolsky}%
  \BibitemOpen
  \bibfield  {author} {\bibinfo {author} {\bibfnamefont {D.}~\bibnamefont
  {Podolsky}}\ and\ \bibinfo {author} {\bibfnamefont {S.}~\bibnamefont
  {Sachdev}},\ }\href@noop {} {\bibfield  {journal} {\bibinfo  {journal} {Phys.
  Rev. B}\ }\textbf {\bibinfo {volume} {86}},\ \bibinfo {pages} {054508}
  (\bibinfo {year} {2012})}\BibitemShut {NoStop}%
\bibitem [{\citenamefont {Endres}\ \emph {et~al.}(2012)\citenamefont {Endres},
  \citenamefont {Fukuhara}, \citenamefont {Pekker}, \citenamefont {Cheneau},
  \citenamefont {Schau$\beta$}, \citenamefont {Gross}, \citenamefont {Demler},
  \citenamefont {Kuhr},\ and\ \citenamefont {Bloch}}]{bloch}%
  \BibitemOpen
  \bibfield  {author} {\bibinfo {author} {\bibfnamefont {M.}~\bibnamefont
  {Endres}}, \bibinfo {author} {\bibfnamefont {T.}~\bibnamefont {Fukuhara}},
  \bibinfo {author} {\bibfnamefont {D.}~\bibnamefont {Pekker}}, \bibinfo
  {author} {\bibfnamefont {M.}~\bibnamefont {Cheneau}}, \bibinfo {author}
  {\bibfnamefont {P.}~\bibnamefont {Schau$\beta$}}, \bibinfo {author}
  {\bibfnamefont {C.}~\bibnamefont {Gross}}, \bibinfo {author} {\bibfnamefont
  {E.}~\bibnamefont {Demler}}, \bibinfo {author} {\bibfnamefont
  {S.}~\bibnamefont {Kuhr}}, \ and\ \bibinfo {author} {\bibfnamefont
  {I.}~\bibnamefont {Bloch}},\ }\href@noop {} {\bibfield  {journal} {\bibinfo
  {journal} {Nature}\ }\textbf {\bibinfo {volume} {487}},\ \bibinfo {pages}
  {454} (\bibinfo {year} {2012})}\BibitemShut {NoStop}%
\bibitem [{\citenamefont {Bradley}\ and\ \citenamefont
  {Doniach}(1984)}]{doniach}%
  \BibitemOpen
  \bibfield  {author} {\bibinfo {author} {\bibfnamefont {R.}~\bibnamefont
  {Bradley}}\ and\ \bibinfo {author} {\bibfnamefont {S.}~\bibnamefont
  {Doniach}},\ }\href@noop {} {\bibfield  {journal} {\bibinfo  {journal} {Phys.
  Rev. B}\ }\textbf {\bibinfo {volume} {30}},\ \bibinfo {pages} {1138}
  (\bibinfo {year} {1984})}\BibitemShut {NoStop}%
\bibitem [{\citenamefont {Giamarchi}(2004)}]{giamarchibook}%
  \BibitemOpen
  \bibfield  {author} {\bibinfo {author} {\bibfnamefont {T.}~\bibnamefont
  {Giamarchi}},\ }\href@noop {} {\emph {\bibinfo {title} {Quantum physics in
  one dimension}}},\ Vol.\ \bibinfo {volume} {121}\ (\bibinfo  {publisher}
  {Oxford university press},\ \bibinfo {year} {2004})\BibitemShut {NoStop}%
\bibitem [{\citenamefont {Bard}\ \emph {et~al.}(2017)\citenamefont {Bard},
  \citenamefont {Protopopov}, \citenamefont {Gornyi}, \citenamefont
  {Shnirman},\ and\ \citenamefont {Mirlin}}]{mirlininsulator}%
  \BibitemOpen
  \bibfield  {author} {\bibinfo {author} {\bibfnamefont {M.}~\bibnamefont
  {Bard}}, \bibinfo {author} {\bibfnamefont {I.}~\bibnamefont {Protopopov}},
  \bibinfo {author} {\bibfnamefont {I.}~\bibnamefont {Gornyi}}, \bibinfo
  {author} {\bibfnamefont {A.}~\bibnamefont {Shnirman}}, \ and\ \bibinfo
  {author} {\bibfnamefont {A.}~\bibnamefont {Mirlin}},\ }\href@noop {}
  {\bibfield  {journal} {\bibinfo  {journal} {Phys. Rev. B}\ }\textbf {\bibinfo
  {volume} {96}},\ \bibinfo {pages} {064514} (\bibinfo {year}
  {2017})}\BibitemShut {NoStop}%
\bibitem [{\citenamefont {Matveev}\ \emph {et~al.}(2002)\citenamefont
  {Matveev}, \citenamefont {Larkin},\ and\ \citenamefont {Glazman}}]{matveev}%
  \BibitemOpen
  \bibfield  {author} {\bibinfo {author} {\bibfnamefont {K.}~\bibnamefont
  {Matveev}}, \bibinfo {author} {\bibfnamefont {A.}~\bibnamefont {Larkin}}, \
  and\ \bibinfo {author} {\bibfnamefont {L.}~\bibnamefont {Glazman}},\
  }\href@noop {} {\bibfield  {journal} {\bibinfo  {journal} {Phys. Rev. Lett.}\
  }\textbf {\bibinfo {volume} {89}},\ \bibinfo {pages} {096802} (\bibinfo
  {year} {2002})}\BibitemShut {NoStop}%
\bibitem [{\citenamefont {Rastelli}\ \emph {et~al.}(2013)\citenamefont
  {Rastelli}, \citenamefont {Pop},\ and\ \citenamefont {Hekking}}]{rastelli}%
  \BibitemOpen
  \bibfield  {author} {\bibinfo {author} {\bibfnamefont {G.}~\bibnamefont
  {Rastelli}}, \bibinfo {author} {\bibfnamefont {I.~M.}\ \bibnamefont {Pop}}, \
  and\ \bibinfo {author} {\bibfnamefont {F.~W.}\ \bibnamefont {Hekking}},\
  }\href@noop {} {\bibfield  {journal} {\bibinfo  {journal} {Phys. Rev. B}\
  }\textbf {\bibinfo {volume} {87}},\ \bibinfo {pages} {174513} (\bibinfo
  {year} {2013})}\BibitemShut {NoStop}%
\bibitem [{\citenamefont {Chow}\ \emph {et~al.}(1998)\citenamefont {Chow},
  \citenamefont {Delsing},\ and\ \citenamefont {Haviland}}]{haviland}%
  \BibitemOpen
  \bibfield  {author} {\bibinfo {author} {\bibfnamefont {E.}~\bibnamefont
  {Chow}}, \bibinfo {author} {\bibfnamefont {P.}~\bibnamefont {Delsing}}, \
  and\ \bibinfo {author} {\bibfnamefont {D.~B.}\ \bibnamefont {Haviland}},\
  }\href@noop {} {\bibfield  {journal} {\bibinfo  {journal} {Phys. Rev. Lett.}\
  }\textbf {\bibinfo {volume} {81}},\ \bibinfo {pages} {204} (\bibinfo {year}
  {1998})}\BibitemShut {NoStop}%
\bibitem [{\citenamefont {Cedergren}\ \emph {et~al.}(2017)\citenamefont
  {Cedergren}, \citenamefont {Ackroyd}, \citenamefont {Kafanov}, \citenamefont
  {Vogt}, \citenamefont {Shnirman},\ and\ \citenamefont {Duty}}]{duty}%
  \BibitemOpen
  \bibfield  {author} {\bibinfo {author} {\bibfnamefont {K.}~\bibnamefont
  {Cedergren}}, \bibinfo {author} {\bibfnamefont {R.}~\bibnamefont {Ackroyd}},
  \bibinfo {author} {\bibfnamefont {S.}~\bibnamefont {Kafanov}}, \bibinfo
  {author} {\bibfnamefont {N.}~\bibnamefont {Vogt}}, \bibinfo {author}
  {\bibfnamefont {A.}~\bibnamefont {Shnirman}}, \ and\ \bibinfo {author}
  {\bibfnamefont {T.}~\bibnamefont {Duty}},\ }\href@noop {} {\bibfield
  {journal} {\bibinfo  {journal} {Phys. Rev. Lett.}\ }\textbf {\bibinfo
  {volume} {119}},\ \bibinfo {pages} {167701} (\bibinfo {year}
  {2017})}\BibitemShut {NoStop}%
\bibitem [{\citenamefont {Kuzmin}\ \emph {et~al.}(2018)\citenamefont {Kuzmin},
  \citenamefont {Mencia}, \citenamefont {Grabon}, \citenamefont {Mehta},
  \citenamefont {Lin},\ and\ \citenamefont {Manucharyan}}]{vlad}%
  \BibitemOpen
  \bibfield  {author} {\bibinfo {author} {\bibfnamefont {R.}~\bibnamefont
  {Kuzmin}}, \bibinfo {author} {\bibfnamefont {R.}~\bibnamefont {Mencia}},
  \bibinfo {author} {\bibfnamefont {N.}~\bibnamefont {Grabon}}, \bibinfo
  {author} {\bibfnamefont {N.}~\bibnamefont {Mehta}}, \bibinfo {author}
  {\bibfnamefont {Y.-H.}\ \bibnamefont {Lin}}, \ and\ \bibinfo {author}
  {\bibfnamefont {V.~E.}\ \bibnamefont {Manucharyan}},\ }\href@noop {}
  {\bibfield  {journal} {\bibinfo  {journal} {arXiv preprint arXiv:1805.07379}\
  } (\bibinfo {year} {2018})}\BibitemShut {NoStop}%
\bibitem [{\citenamefont {Luther}\ and\ \citenamefont {Emery}(1974)}]{luther}%
  \BibitemOpen
  \bibfield  {author} {\bibinfo {author} {\bibfnamefont {A.}~\bibnamefont
  {Luther}}\ and\ \bibinfo {author} {\bibfnamefont {V.~J.}\ \bibnamefont
  {Emery}},\ }\href {\doibase 10.1103/PhysRevLett.33.589} {\bibfield  {journal}
  {\bibinfo  {journal} {Phys. Rev. Lett.}\ }\textbf {\bibinfo {volume} {33}},\
  \bibinfo {pages} {589} (\bibinfo {year} {1974})}\BibitemShut {NoStop}%
\bibitem [{\citenamefont {Zamolodchikov}(1995)}]{zamolodchikovmass}%
  \BibitemOpen
  \bibfield  {author} {\bibinfo {author} {\bibfnamefont {A.~B.}\ \bibnamefont
  {Zamolodchikov}},\ }\href@noop {} {\bibfield  {journal} {\bibinfo  {journal}
  {International Journal of Modern Physics A}\ }\textbf {\bibinfo {volume}
  {10}},\ \bibinfo {pages} {1125} (\bibinfo {year} {1995})}\BibitemShut
  {NoStop}%
\bibitem [{\citenamefont {Coleman}(1975)}]{coleman}%
  \BibitemOpen
  \bibfield  {author} {\bibinfo {author} {\bibfnamefont {S.}~\bibnamefont
  {Coleman}},\ }\href {\doibase 10.1103/PhysRevD.11.2088} {\bibfield  {journal}
  {\bibinfo  {journal} {Phys. Rev. D}\ }\textbf {\bibinfo {volume} {11}},\
  \bibinfo {pages} {2088} (\bibinfo {year} {1975})}\BibitemShut {NoStop}%
\bibitem [{\citenamefont {Baier}\ and\ \citenamefont
  {Pilon}(1991)}]{baier1991axial}%
  \BibitemOpen
  \bibfield  {author} {\bibinfo {author} {\bibfnamefont {R.}~\bibnamefont
  {Baier}}\ and\ \bibinfo {author} {\bibfnamefont {E.}~\bibnamefont {Pilon}},\
  }\href@noop {} {\bibfield  {journal} {\bibinfo  {journal} {Zeitschrift
  f{\"u}r Physik C Particles and Fields}\ }\textbf {\bibinfo {volume} {52}},\
  \bibinfo {pages} {339} (\bibinfo {year} {1991})}\BibitemShut {NoStop}%
\bibitem [{\citenamefont {Mandelstam}(1975)}]{mandelstam}%
  \BibitemOpen
  \bibfield  {author} {\bibinfo {author} {\bibfnamefont {S.}~\bibnamefont
  {Mandelstam}},\ }\href {\doibase 10.1103/PhysRevD.11.3026} {\bibfield
  {journal} {\bibinfo  {journal} {Phys. Rev. D}\ }\textbf {\bibinfo {volume}
  {11}},\ \bibinfo {pages} {3026} (\bibinfo {year} {1975})}\BibitemShut
  {NoStop}%
\bibitem [{\citenamefont {Babujian}\ \emph {et~al.}(1999)\citenamefont
  {Babujian}, \citenamefont {Fring}, \citenamefont {Karowski},\ and\
  \citenamefont {Zapletal}}]{babujian}%
  \BibitemOpen
  \bibfield  {author} {\bibinfo {author} {\bibfnamefont {H.}~\bibnamefont
  {Babujian}}, \bibinfo {author} {\bibfnamefont {A.}~\bibnamefont {Fring}},
  \bibinfo {author} {\bibfnamefont {M.}~\bibnamefont {Karowski}}, \ and\
  \bibinfo {author} {\bibfnamefont {A.}~\bibnamefont {Zapletal}},\ }\href@noop
  {} {\bibfield  {journal} {\bibinfo  {journal} {Nuclear Physics B}\ }\textbf
  {\bibinfo {volume} {538}},\ \bibinfo {pages} {535} (\bibinfo {year}
  {1999})}\BibitemShut {NoStop}%
\bibitem [{\citenamefont {Giamarchi}\ and\ \citenamefont
  {Schulz}(1988)}]{giamarchischulz}%
  \BibitemOpen
  \bibfield  {author} {\bibinfo {author} {\bibfnamefont {T.}~\bibnamefont
  {Giamarchi}}\ and\ \bibinfo {author} {\bibfnamefont {H.}~\bibnamefont
  {Schulz}},\ }\href@noop {} {\bibfield  {journal} {\bibinfo  {journal} {Phys.
  Rev. B}\ }\textbf {\bibinfo {volume} {37}},\ \bibinfo {pages} {325} (\bibinfo
  {year} {1988})}\BibitemShut {NoStop}%
\bibitem [{\citenamefont {Choi}\ \emph {et~al.}(1998)\citenamefont {Choi},
  \citenamefont {Yi}, \citenamefont {Choi}, \citenamefont {Choi},\ and\
  \citenamefont {Lee}}]{choi1998quantum}%
  \BibitemOpen
  \bibfield  {author} {\bibinfo {author} {\bibfnamefont {M.-S.}\ \bibnamefont
  {Choi}}, \bibinfo {author} {\bibfnamefont {J.}~\bibnamefont {Yi}}, \bibinfo
  {author} {\bibfnamefont {M.}~\bibnamefont {Choi}}, \bibinfo {author}
  {\bibfnamefont {J.}~\bibnamefont {Choi}}, \ and\ \bibinfo {author}
  {\bibfnamefont {S.-I.}\ \bibnamefont {Lee}},\ }\href@noop {} {\bibfield
  {journal} {\bibinfo  {journal} {Phys. Rev. B}\ }\textbf {\bibinfo {volume}
  {57}},\ \bibinfo {pages} {R716} (\bibinfo {year} {1998})}\BibitemShut
  {NoStop}%
\bibitem [{\citenamefont {B{\"u}chler}\ \emph {et~al.}(2003)\citenamefont
  {B{\"u}chler}, \citenamefont {Blatter},\ and\ \citenamefont
  {Zwerger}}]{zwerger}%
  \BibitemOpen
  \bibfield  {author} {\bibinfo {author} {\bibfnamefont {H.}~\bibnamefont
  {B{\"u}chler}}, \bibinfo {author} {\bibfnamefont {G.}~\bibnamefont
  {Blatter}}, \ and\ \bibinfo {author} {\bibfnamefont {W.}~\bibnamefont
  {Zwerger}},\ }\href@noop {} {\bibfield  {journal} {\bibinfo  {journal} {Phys.
  Rev. Lett.}\ }\textbf {\bibinfo {volume} {90}},\ \bibinfo {pages} {130401}
  (\bibinfo {year} {2003})}\BibitemShut {NoStop}%
\bibitem [{\citenamefont {Bard}\ \emph {et~al.}(2018)\citenamefont {Bard},
  \citenamefont {Protopopov},\ and\ \citenamefont {Mirlin}}]{mirlindecay}%
  \BibitemOpen
  \bibfield  {author} {\bibinfo {author} {\bibfnamefont {M.}~\bibnamefont
  {Bard}}, \bibinfo {author} {\bibfnamefont {I.}~\bibnamefont {Protopopov}}, \
  and\ \bibinfo {author} {\bibfnamefont {A.}~\bibnamefont {Mirlin}},\
  }\href@noop {} {\bibfield  {journal} {\bibinfo  {journal} {Phys. Rev. B}\
  }\textbf {\bibinfo {volume} {98}},\ \bibinfo {pages} {224513} (\bibinfo
  {year} {2018})}\BibitemShut {NoStop}%
\bibitem [{\citenamefont {Houzet}\ and\ \citenamefont
  {Glazman}(2019)}]{HouzetGlazman}%
  \BibitemOpen
  \bibfield  {author} {\bibinfo {author} {\bibfnamefont {M.}~\bibnamefont
  {Houzet}}\ and\ \bibinfo {author} {\bibfnamefont {L.~I.}\ \bibnamefont
  {Glazman}},\ }\href@noop {} {\bibfield  {journal} {\bibinfo  {journal} {arXiv
  preprint arXiv:1901.01515}\ } (\bibinfo {year} {2019})}\BibitemShut {NoStop}%
\end{thebibliography}%

\pagebreak

\appendix 
\renewcommand{\thefigure}{A\arabic{figure}}

\setcounter{figure}{0}

\section{Formal sine-Gordon representation of phase slips}

The JJ chain Hamiltonian is written as
\begin{align}
H = \sum_{i,j}&U_{i,j}(N_i-Q_i)(N_j-Q_j)\nonumber\\ &+\sum_i E_J[1-\cos(\theta_i-\theta_{i-1})],
\label{eq:HJJC}
\end{align}
where $N_i$ is number of Cooper pairs on site $i$, with strayed charge $Q_i$ that represents disorder, $E_J$ is the Josephson energy of the junctions and $\theta_i$ is phase of site $i$. $U_{i,j}$ is the charging energy between $i$ and $j$.
Starting with the canonical commutation relations for charge and phase 
\begin{align}
&[\theta_m,N_n]=i \delta_{mn},
\end{align}
one can define a new variable
\begin{align}
&{\phi}_n=\sum_m K(m-n)N_m\\
& K(n)=\pi[1-\tanh{\alpha (n-1/2)}]/2, 
\end{align}
\begin{figure}
\centering
	\includegraphics[width=0.8\textwidth,angle=270]{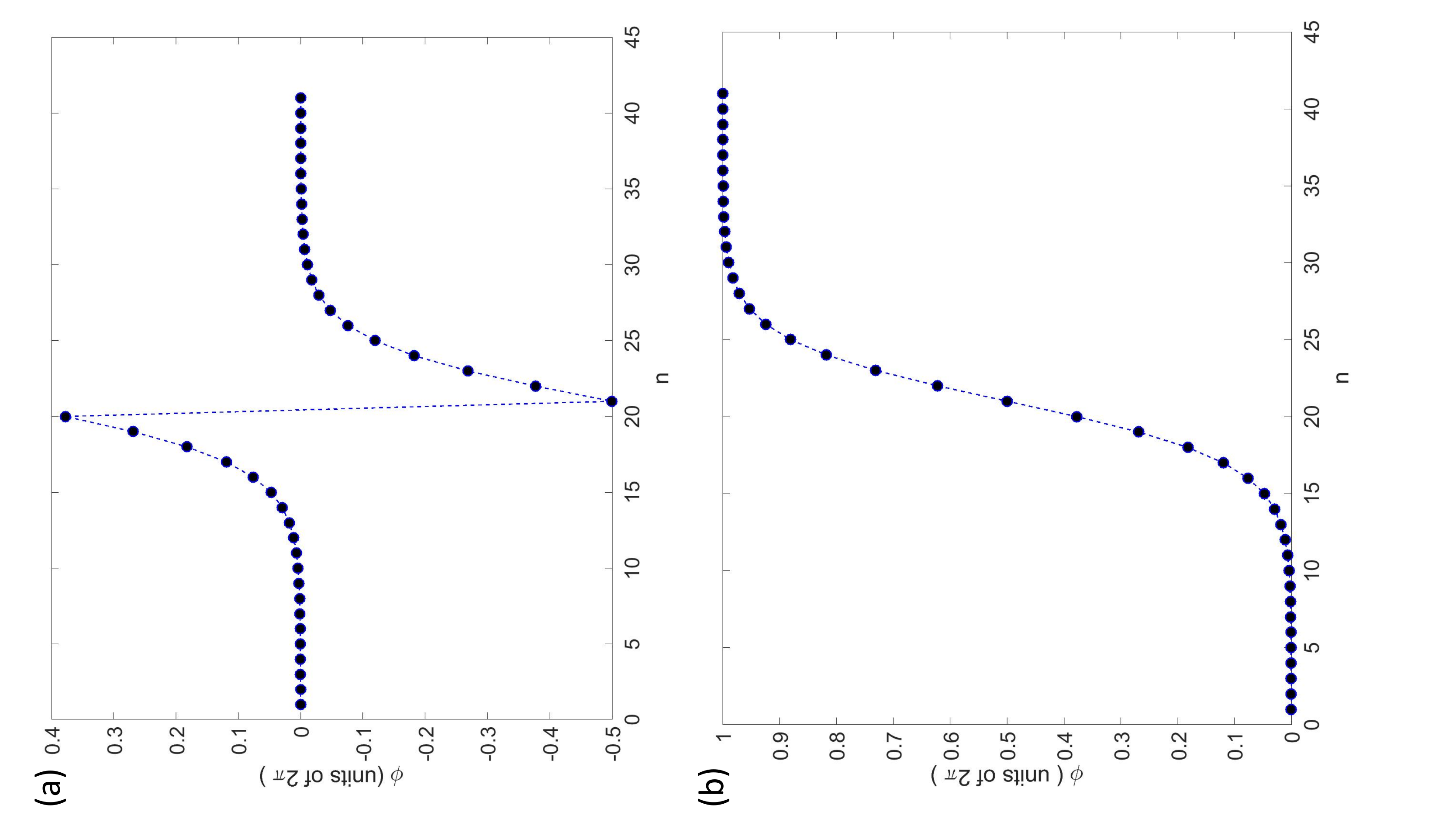}
	\caption{(a) Phase profile $\phi_n$ for the microscopic JJ chain model. The phase-slip phase profile is discontinuous but 
local in the sense that it vanishes away from the phase slip.
(b) The discontinuity can be removed at the expense of introducing non-locality to produce an unwound phase-slip.
} 
\label{Figphaseslip}
\end{figure}
where $\alpha$ represents the width of the phase slip profile. The new variable ${\phi}$ satisfies the commutation relation 
\begin{align}
&[\theta_m,{\phi}_n]=i K(m-n).
\end{align}
Using the Campbell-Baker-Hausdorf relation, 
\begin{align}
&e^{-2i{\phi}_n}\theta_m e^{2i{\phi}_n}=\theta_m-2K(m-n),
\end{align}
one can show that $e^{2i{\phi}_n}$ creates a phase-slip 
with the profile shown in Fig.~\ref{Figphaseslip}.
Thus, the phase slip is generated by the sine term of the sine-Gordon model
\begin{align}
&H_{sine}=g\sum_i\cos{2\phi_i}.
\end{align}

In order to properly include the variable $\theta$, we need to transform to dual variables, which are 
defined by the momentum 
\begin{align}
&\Pi_m=(\theta_{m+1}-\theta_m)/\pi,
\end{align}
which is canonically conjugate to $\theta_n$ as 
\begin{align}
&[\phi_n,\Pi_m]=i[K(m+1-n)-K(m-n)]\approx i\delta_{nm}
\end{align}
at length scale much larger than the width of the profile. In such limit, the JJ chain Hamiltonian, in terms of $\Pi$ and $\theta$, becomes 
\begin{align}
H = &\sum_{i,j}U_{i,j}(\frac{\phi_i-\phi_{i-1}}{\pi}-Q_i)(\frac{\phi_j-\phi_{j-1}}{\pi}-Q_j)\nonumber\\ &+\sum_i E_J(1-\cos\pi\Pi_j)
\end{align}
In the large $E_J$ regime, one can expand the Josephson energy to quadratic term. In this case, we should also include the phase slip term $H_{sine}$. This gives
\begin{align}
H = &\sum_{i,j}U_{i,j}(\frac{\phi_i-\phi_{i-1}}{\pi}-Q_i)(\frac{\phi_j-\phi_{j-1}}{\pi}-Q_j)\nonumber\\ &+\frac{E_J}{2}\sum_i (\pi\Pi_j)^2 + g\sum_i\cos 2\phi_i
\end{align}
Next we transform $\phi_i\rightarrow \phi_i+\pi\sum_{j\leq i}Q_j$, this gives rise to the Aharonov-Casher phase in the phase slip term,
\begin{align}
H = &\sum_{i,j}\frac{U_{i,j}}{\pi^2}(\phi_i-\phi_{i-1})(\phi_j-\phi_{j-1})\nonumber\\ &+\frac{E_J}{2}\sum_i (\pi\Pi_j)^2 + g\sum_i\cos (2\phi_i+2\pi\sum_{j\leq i}Q_j)
\end{align}
Finally, in the continuum limit ($\phi\rightarrow \phi(x)$, $\Pi\rightarrow j(x)$), considering length scale to be larger than the Coulomb interaction range $\sqrt{E_0/E_1}$ where $U_{i,j}\approx 4E_0\delta_{i,j}$, the Hamiltonian becomes sine-Gordon-typed,
\begin{align}
H = \int dx [\frac{\pi^2E_J}{2}j(x)^2 &+\frac{4E_0}{\pi^2}\partial\phi(x)^2 \nonumber\\ &+ g\cos(2\phi(x)+2\pi \int_{x'<x} Q(x'))].
\end{align}

\section{Phase-slips from microscopic model}

In this appendix, we show that the partition function of the sine-Gordon model matches that of the microscopic JJ chain model under certain limit with certain value of $g$. This in principle allows one to determine $g$. For simplicity, periodic boundary condition is chosen for both JJ chain and sine-Gordon model.

We start with the microscopic Hamiltonian of the JJ, which is given by
\begin{align}
H = \sum_{i,j} N_i U_{ij} N_j + E_J\sum_i (1-\cos\Delta\theta_i)
\end{align}
where $\Delta \theta_i = \theta_{i} - \theta_{i-1}$. This gives the imaginary time action
\begin{align}
S[\theta] = \int  \frac{1}{4}\sum_{i,j}\dot{\theta}_i C_{i,j} \dot{\theta}_j +E_J\sum_i (1-\cos\Delta\theta_i) d\tau,
\end{align}
where $C\equiv U^{-1}$ is the capacitance matrix.  The partition function $Z$ in path integral form is
\begin{align}
Z_{JJC} = \oint \mathcal{D}[\theta] e^{-S[\theta]}
\end{align}
where the integration limit of $S$ is from $0$ to $\beta$. Due to the $2\pi$ periodicity in $\theta$, the configurations can have B.C's with $2\pi n$ jumps in $\theta_{N+1} -\theta_{1}$ and $\int d\tau \partial_\tau \theta_i$. This is equivalent to including particular kinds of branch cuts into the configurations and restoring periodic boundary conditions on both time and position. 

The branch cuts transforms $(\partial_\tau, \Delta) \rightarrow (\partial_\tau - a_0, \Delta - a_1)$  with a vector potential $\mathbf{a} = (a_0, a_1)$,
\begin{align}
Z_{JJC} = \sum_\mathbf{a}\int \mathcal{D}[\theta]e^{-S_\mathbf{a}[\theta]} 
\end{align}
and the $\mathbf{a}$-dependent action $S_\mathbf{a}$ is
\begin{align}
S_{\mathbf{a}}[\theta] = \int d\tau [\sum_{i,j}(\dot{\theta}_i-a_{i0}) \frac{C_{i,j}}{4} (\dot{\theta}_j-a_{j0})\nonumber\\ +E_J\sum_i (1-\cos(\Delta\theta_i- a_{i1}))]
\end{align}
where $\mathbf{a}$ labels the sets of branch cuts. Since the path integral is invariant under a change of variable $\theta \rightarrow \theta + \Lambda$, $(a_{0},a_{1}) \rightarrow (a_{0} + \partial_\tau \Lambda, a_1 + \Delta \Lambda)$ is a gauge symmetry. Therefore, we can choose the non gauge equivalent branch cuts to be composed of vertical cuts,
\begin{align}
\mathbf{a}_i(\tau) = (0,\pm 2  \pi\delta_{i,x_v} \Theta(\tau_v-\tau))
\label{eq:bc1}
\end{align} 
where $x_v$ and $\tau_v$ are the coordinates of the end point of the branch cuts (vortices/antivortices), which are determined by the ($\pm$) sign; and horizontal branch cuts,
\begin{align}
\mathbf{a}_i(\tau) = (\pm 2\pi\delta(\tau) \Theta(x_{v_1}<i<x_{v_2}),0).
\label{eq:bc2}
\end{align}
Finally, there are also full branch cuts that does not have vortices. These are defined as $\mathbf{a}_i(\tau) = (0, \pm2\pi\delta_{i,0})$ or $\mathbf{a}_i(\tau) = (\pm 2\pi\delta(\tau),0)$.

Next, we split the field into smooth and fast part.
\begin{align}
\theta = \theta_s + \theta_f
\end{align}
where $\theta_s$ is defined by $\sqrt{(\partial_\tau\theta_s-a_0)^2+(\Delta \theta_s-a_1)^2}$ being small.
The configurations that contribute more are ones with $\theta_f \simeq 0$ everywhere except near the vortices ($(x,\tau)\sim (x_v,\tau_v)$), where $\mathcal{L_{\mathbf{a}}}(\dot{\theta}_f,\theta_f)\gg \mathcal{L_{\mathbf{a}}}(\dot{\theta}_s,\theta_s)$. Therefore,
\begin{align}
Z_{JJC} &= \sum_\mathbf{a}\int \mathcal{D}[\theta_s]\mathcal{D}[\theta_f]e^{-S_\mathbf{a}[\theta_s+\theta_f]} \nonumber\\
&\simeq \sum_\mathbf{a}\int \mathcal{D}[\theta_s]e^{-S_\mathbf{a}[\theta_s]} \int \mathcal{D}[\theta_f]e^{-S_\mathbf{a}[\theta_f]} \nonumber\\
&\simeq \sum_\mathbf{a}\int \mathcal{D}[\theta_s]e^{-S_\mathbf{a}[\theta_s]} \prod_{v\in\mathbf{a}}\int \mathcal{D}[\theta_f]e^{-S_v[\theta_f]} \nonumber\\
&\equiv \sum_\mathbf{a} \gamma^{n_\mathbf{a}}\int \mathcal{D}[\theta_s]e^{-S_\mathbf{a}[\theta_s]}
\end{align}
where $v$ labels vortices in the set of branch cuts $\mathbf{a}$, $S_v[\theta_f] = \sum_{x\sim x_v}\int_{\tau\sim\tau_v} d\tau\mathcal{L}(\dot{\theta}_f,\theta_f)$ and $\gamma$ is the contribution from integrating $\theta_f$ near a single vortex/antivortex, $n_\mathbf{v}$ is the number of vortices. Finally, since $\theta_s$ is smooth, we can approximate $S_\mathbf{a}[\theta_s]$ by the free action $S_{0,\mathbf{a}}[\theta_s] = \int d\tau [\sum_{i,j}(\dot{\theta}_i-a_{i0}) \frac{C_{i,j}}{4} (\dot{\theta}_j-a_{j0}) +\frac{E_J}{2}\sum_i (\Delta\theta_i- a_{i1})^2]$, this gives the following form of the partition function.
\begin{align}
Z_{JJC} = \sum_\mathbf{a} \gamma^{n_\mathbf{a}}\int \mathcal{D}[\theta_s]e^{-S_{0,\mathbf{a}}[\theta_s]}
\end{align}
To solve for $Z_{JJC}$, it is more convenient to work in the Fourier space,
\begin{align}
\theta_m(\tau) &= \frac{1}{\sqrt{M}}\sum_{k,\omega} \tilde{\theta}_{k,\omega}e^{i(km-\omega\tau)}\nonumber\\
\mathbf{a}_m(\theta) &= \frac{1}{\sqrt{M}}\sum_{k,\omega}  \tilde{\mathbf{a}}_{k,\omega}e^{i(km-\omega\tau)},
\end{align}
where $M$ is the system size and $k\in \{\frac{2n\pi}{M}\vert n \in\{-M/2... M/2\}\}$, $\omega\in \{\frac{2n\pi}{\beta}\vert n\in \mathbb{Z}\}$. This gives
\begin{align}
&S_{0,\mathbf{a}}[\theta_s]/\beta= \nonumber\\&  \sum_{(k,\omega)\neq (0,0)} A_{k,\omega}|\tilde{\theta}_{k,\omega}-\frac{(E_J/2)(1-e^{ik})\tilde{a}_1+i(\omega \tilde{C}_k /4) \tilde{a}_0}{A_{k,\omega}}|^2 \nonumber\\
&+\frac{\tilde{C}_0}{4}|\tilde{a}_{0;(0,0)}|^2+\frac{E_J}{2}|\tilde{a}_{1;(0,0)}|^2 \nonumber\\&+(\frac{E_J \tilde{C}_0}{8})\sum_{(k,\omega)\neq (0,0)}\frac{|(1-e^{-ik})\tilde{a}_0+i\omega \tilde{a}_1|^2}{A_{k,\omega}}
\end{align}
where $\tilde{C}_k = \sum_m C_{m} e^{ikm}$, $A_{k,\omega} = \omega^2\tilde{C}_k/4+2E_J\sin^2(k/2)$. 
Next, using $\mathcal{D}[\theta_s] = J \prod_{k,\omega} d\tilde{\theta}_{k,\omega}$,
\begin{align}
&Z_{JJC} = J\sum\limits_{\mathbf{a}} \{\gamma^{n_\mathbf{a}} e^{- \beta[(\tilde{C}_0/4)|\tilde{a}_{0;(0,0)}|^2+(E_J/2)|\tilde{a}_{1;(0,0)}|^2]}\nonumber\\ &\quad\times \prod_{(k,\omega)\neq(0,0)} [e^{-\frac{\beta E_J\tilde{C}_k}{8A_{k,\omega}}|(1-e^{-ik})\tilde{a}_0+i\omega \tilde{a}_1|^2}\nonumber\\ &\qquad\qquad\qquad\qquad\qquad\times\int  d\tilde{\theta}_{k,\omega}e^{-\beta  A_{k,\omega}|\tilde{\theta}_{k,\omega}|^2}]\} \nonumber\\
&\propto \sum\limits_{\mathbf{a}}\{ \gamma^{n_\mathbf{a}} e^{- \beta[(\tilde{C}_0/4)|\tilde{a}_{0;(0,0)}|^2+(E_J/2)|\tilde{a}_{1;(0,0)}|^2]} \nonumber\\ &\qquad\qquad\qquad\times\prod_{(k,\omega)\neq(0,0)} e^{-\frac{\beta E_J\tilde{C}_k}{8A_{k,\omega}}|(1-e^{-ik})\tilde{a}_0+i\omega \tilde{a}_1|^2}\}
\end{align}
Note that the set of broken branch cuts $\mathbf{a}$ and the vortex configuration has a 1-to-1 correspondence. Therefore, we can relabel $\mathbf{a}$ by $(n_\tau,n_x,\mathbf{v})$ which denotes the number of $2\pi$ jumps at the $x$ and $\tau$ boundary and the vortex configuration. 

For the second term, by Eq.~\ref{eq:bc1} and Eq.~\ref{eq:bc2}, $\tilde{\mathbf{a}}_{(0,0)}$ are related to the polarization of $\mathbf{v}$ up to full branch cuts, 
\begin{align}
\tilde{a}_{0;(0,0)} &= \frac{1}{\beta\sqrt{M}}\sum_{m}\int a_{0,m}(\tau) d\tau \nonumber\\ &= \frac{2\pi}{\beta\sqrt{M}}(M n_\tau + \sum_{v\in\mathbf{v}} x_v s_v)\nonumber\\ &\equiv \frac{2\pi}{\beta\sqrt{M}}(M n_\tau+P_{x,\mathbf{v}})
\end{align}
\begin{align}
\tilde{a}_{1;(0,0)} &= \frac{1}{\beta\sqrt{M}}\sum_{m}\int a_{1,m}(\tau) d\tau \nonumber\\ &= \frac{2\pi}{\beta\sqrt{M}}(\beta n_x+\sum_{v\in\mathbf{v}} \tau_v s_v)\nonumber\\ &\equiv \frac{2\pi}{\beta\sqrt{M}}(\beta n_x+P_{\tau,\mathbf{v}})
\end{align}
where $s_v=\pm 1$ for vortices/antivortices. The polarization is defined as $P_\mathbf{v} = (\sum_{v\in \mathbf{v}} \tau_v s_v, \sum_{v\in\mathbf{v}}x_vs_v)$. Next, for the third term, the exponent contains a discrete version of $curl\: \mathbf{a}$ which can also be rewritten as the vortex density,
\begin{align}
|(1-e^{-ik})\tilde{a}_0+i\omega\tilde{a}_1|^2 = \frac{4\pi^2}{\beta^2 M}|B_{\mathbf{v};k,\omega}|^2
\end{align}
where $B_{\mathbf{v};k,\omega}\equiv \sum_{v\in\mathbf{v}}s_ve^{-i(kx_v-\omega\tau_v)}$.
Therefore, the partition function represented in terms of the vortex configuration is
\begin{align}
Z_{JJC} \sim &\sum_{\mathbf{v}}\{\gamma^{n_\mathbf{v}}\sum_{n_x,n_\tau}e^{-[\frac{\pi^2\tilde{C}_0}{\beta M}(Mn_\tau+P_{x,\mathbf{v}})^2+\frac{2\pi^2 E_J}{\beta M}(\beta n_x + P_{\tau,\mathbf{v}})^2]}\nonumber\\ &\times\prod_{(k,\omega)\neq(0,0)}e^{-\frac{\pi^2E_J\tilde{C}_k}{2\beta M}(\frac{|B_{\mathbf{v};k,\omega}|^2}{A_{k,\omega}})}\}
\end{align}

Finally, to make connection to the s-G model that is going to be introduced later, we perform a Poisson resummation on the discrete Gaussians. This gives
\begin{align}
Z_{JJC} \sim &\sum_{\mathbf{v}}\{\gamma^{n_\mathbf{v}}\sum_{n}e^{-\frac{\beta}{M \tilde{C}_0}n^2 + i\frac{2\pi P_{x,\mathbf{v}}}{M} n}\sum_{n}e^{-\frac{M}{2\beta E_J}n^2 + i\frac{2\pi P_{\tau,\mathbf{v}}}{M} n}\nonumber\\&\times\prod_{(k,\omega)\neq(0,0)}e^{-\frac{\pi^2E_J\tilde{C}_k}{2\beta M}(\frac{|B_{\mathbf{v};k,\omega}|^2}{A_{k,\omega}})}\}
\label{eq:Z_JJC}
\end{align}

Next, we derive the partition function $Z_{s-G}$ for sine-Gordon model defined as
\begin{align}
H_{s-G} = \sum_{i,j}N_j U_{i,j} N_j +\frac{E_J}{2}\sum_j (\theta_j-\theta_{j-1})^2 + g \sum_j \cos(2\phi_j).
\end{align}
It is convenient to perform a charge-vortex transformation $(\theta, N)\rightarrow(\phi,\Pi)$, with the relation
\begin{align}
N_i &= (\theta_i-\theta_{i-1})/\pi \nonumber\\
\Pi_i &= (\phi_{i+1}-\phi_i)/\pi.
\end{align}
One can easily check that the commutation relation $[\phi_i, \Pi_j] = i\delta_{i,j}$ is preserved. In terms of $\phi$ and $\Pi$,
\begin{align}
H_{s-G} =  &\sum_{i,j}(\phi_i-\phi_{i-1}) \frac{U_{i,j}}{\pi^2} (\phi_j-\phi_{j-1}) \nonumber
\\&+ \frac{\pi^2 E_J}{2}\sum_j \Pi_j^2 + g \sum_j cos(2\phi_j).
\end{align}
Using the path integral formalism, the partition function can be written as,
\begin{align}
Z_{s-G} = \sum_{n_x,n_\tau}\int_{\phi(\beta) = \phi(0)}\mathcal{D}[\phi] e^{-S_{n_x,n_\tau}}
\end{align}
where $n_x,n_\tau$ labels the boundary condition defined by
\begin{align}
&2\phi_{M+1}(\tau) = 2\phi_1(\tau) + 2\pi n_x \nonumber\\
&2\phi_{j}(\beta) = 2\phi_{j}(0) + 2\pi n_\tau
\end{align}
and the imaginary time action can be written as
\begin{align}
S_{n_x,n_\tau} =& \int^\beta_0 d\tau [\frac{1}{2\pi^2 E_J}\sum_j (\dot{\phi}_j+\frac{\pi n_\tau}{\beta})^2  \nonumber\\ &+\sum_{i,j}(\phi_i-\phi_{i-1}+\frac{\pi n_x}{M}) \frac{U_{i,j}}{\pi^2} (\phi_j-\phi_{j-1}+\frac{\pi n_x}{M})\nonumber\\ &+ g \sum_j \cos (2\phi_j+\frac{2\pi j}{M}n_x + \frac{2\pi \tau}{\beta}n_\tau)].
\end{align}
Working perturbatively in small $g$,
\begin{align}
Z_{s-G}=\sum_{n_x,n_\tau}\sum\limits_n \frac{1}{n!}(-\frac{g}{2})^n  \sum_{l_1,l_2,...,l_n} \int \prod_{j=1}^n d\tau_j \sum_{\{\mathbf{s}\}}\nonumber\\ \int\mathcal{D}[\phi] e^{-S_{0,n_x,n_\tau} +i \sum_js_j[2\phi_{l_j}(\tau_j)+\frac{2\pi l_j}{M}n_x + \frac{2\pi\tau_j}{\beta}n_\tau]}
\end{align}
where $\{\mathbf{s}\}$ sums over the signs ($\pm 1$) and the free action $S_0$ is given by
\begin{align}
S_{0;n_x,n_\tau}&= \int^\beta_0 d\tau [\frac{1}{2\pi^2 E_J}\sum_j (\dot{\phi}_j+\frac{\pi n_\tau}{\beta})^2  \nonumber\\ &+\sum_{i,j}(\phi_i-\phi_{i-1}+\frac{\pi n_x}{M}) \frac{U_{i,j}}{\pi^2} (\phi_j-\phi_{j-1}+\frac{\pi n_x}{M})\nonumber\\
&= \frac{M}{2\beta E_J}n_\tau^2+\frac{\beta \tilde{U}_0}{M}n_x^2 \nonumber\\&+\sum_{(k,\omega)\neq (0,0)}[\frac{\beta }{2\pi^2 E_J} \omega^2+\frac{4\beta\sin^2 (k/2)}{\pi^2}\tilde{U}_{k}]\vert \tilde{\phi}_{k,\omega}\vert^2.
\end{align}
where $\tilde{U}_{k} = \sum_l U_le^{ikl}$ and the Fourier transformation of the $\phi$ fields are defined as
\begin{align}
&\phi_l(\tau) = \frac{1}{\sqrt{M}}\sum_{k,\omega}\tilde{\phi}_{k,\omega}e^{i(kl-\omega\tau)} \nonumber\\
&\tilde{\phi}_{k,\omega} = \frac{1}{\beta \sqrt{M}} \sum_l\int d\tau \phi_l(\tau)e^{-i(kl-\omega\tau)}
\end{align}
where $M$, $k$, and $\omega$ has the same definition as above. Let $\mathcal{D}[\phi] = J \prod_{k,\omega} d\tilde{\phi}_{k,\omega}$,
\begin{align}
Z_{s-G} &= J\sum\limits_{\mathbf{v}} (-\frac{g}{2})^{n_\mathbf{v}}\nonumber\\&\times\sum_{n_\tau}e^{-\frac{M}{2\beta E_J}n_\tau^2 + i\frac{2\pi P_{\tau,\mathbf{v}}}{\beta}n_\tau}\sum_{n_x}e^{-\frac{\beta\tilde{U}_0}{M}n_x^2 + i\frac{2\pi P_{x,\mathbf{v}}}{M}n_x}\nonumber\\&\times \prod_{k,\omega} \int d\tilde{\phi}_{k,\omega} e^{-\frac{2\beta\tilde{U}_k}{\pi^2E_J}A_{k,\omega}\vert \tilde{\phi}_{k,\omega}\vert^2 +i \frac{2}{\sqrt{M}}B_{\mathbf{v};k,\omega}\tilde{\phi}_{k,\omega}}
\end{align}
where $\sum_{\mathbf{v}} \equiv \sum\limits_n \frac{1}{n!} \sum_{l_1,l_2,...,l_n} \int \prod_j^n d\tau_j \sum_{\{\mathbf{s}\}}$ is an analog of the sum over vortex configurations and $n_{\mathbf{v}}$ represents the number of vortices. We will use this notation and index the vortices by $v$. Again, $A_{k,\omega} = \omega^2/4\tilde{U}_k+2E_J\sin^2 (k/2)$, $B_{\mathbf{v};k,\omega} = \sum_{v\in\mathbf{v}} s_ve^{i(kx_v-\omega \tau_v)}$ and $P_{\mathbf{v}}$ is the polarization of the charges with the same definition, $(P_{\tau,\mathbf{v}},P_{x,\mathbf{v}})\equiv (\sum_{v\in\mathbf{v}} \tau_v s_v, \sum_{v\in\mathbf{v}} x_vs_v)$. First, the integral over $\tilde{\phi}'s$ are Gaussian integrals
\begin{align}
&\prod_{k,\omega} \int d\tilde{\phi}_{k,\omega} e^{-\frac{2\beta\tilde{U}_k}{\pi^2E_J}A_{k,\omega}\vert \tilde{\phi}_{k,\omega}\vert^2 +i \frac{2}{\sqrt{M}}B_{\mathbf{v};k,\omega}\tilde{\phi}_{k,\omega}} \nonumber\\
& = \prod_{k,\omega} \sqrt{{\frac{4\pi^3E_J}{\beta\tilde{U}_kA_{k,\omega}}}}e^{-\frac{\pi^2E_J}{2\beta M \tilde{U}_k}(\frac{\vert B_{\mathbf{v};k,\omega}\vert^2}{A_{k,\omega}})}\nonumber\\
& \propto \prod_{k,\omega} e^{-\frac{\pi^2E_J}{2\beta M \tilde{U}_k}(\frac{\vert B_{\mathbf{v};k,\omega}\vert^2}{A_{k,\omega}})}
\end{align}
Using the fact that the only configurations that are going to contribute are the charge neutral ones with $\sum_{v\in\mathbf{v}}s_v = 0$ and $n_\mathbf{v}$ is thereby an even number,
\begin{align}
Z_{s-G} &\sim \sum\limits_{\mathbf{v}} (\frac{g}{2})^{n_\mathbf{v}} \nonumber\\ &\times\sum_{n}e^{-\frac{M}{2\beta E_J}n^2 + i\frac{2\pi P_{\tau,\mathbf{v}}}{\beta}n}\sum_{n}e^{-\frac{\beta\tilde{U}_0}{M}^2 + i\frac{2\pi P_{x,\mathbf{v}}}{M}n} \nonumber\\&\times\prod_{k,\omega} e^{-\frac{\pi^2E_J}{2\beta M \tilde{U}_k}(\frac{\vert B_{\mathbf{v};k,\omega}\vert^2}{A_{k,\omega}})}.
\end{align}
One can see that this form exactly matches $Z_{JJC}$ in (\ref{eq:Z_JJC}) with $g$ chosen as
\begin{align}
g= 2\gamma\equiv 2\int \mathcal{D}[\phi_f]e^{-\sum_{x\sim x_v}\int_{\tau\sim\tau_v} d\tau\mathcal{L}_v(\dot{\phi}_f,\phi_f)}.
\end{align}
 
\section{Massive Dirac model with charge disorder}
Let us consider the Dirac equation 
\begin{align}
i\sigma_z\partial_x\psi+[m\sigma_x-\mu(x)]\psi=0.
\end{align}
This equation is written as 
\begin{align}
\partial_x\psi+[m\sigma_y-i\mu(x)\sigma_z]\psi=0.
\end{align}
Transforming $\psi\rightarrow e^{i\phi(x)\sigma_z}\psi$, where $\phi(x) =\int^x_{x_1} dx' \mu(x') dx'$,
the equation becomes 
\begin{align}
\partial_x\psi=-m[\cos{\phi(x)}\sigma_y-\sin{\phi(x)}\sigma_x]\psi.
\end{align}
Considering the evolution of $\psi(x)$ from $x=x_1$ to $x=x_2$
\begin{align}
\psi(x_2) &= e^{-i\phi(x_2)\sigma_z} \nonumber\\&\times[1-m\int dx' \{\cos\phi(x')\sigma_y-\sin\phi(x')\sigma_x\}]\psi(x_1)\\
&\equiv e^{-i\phi(x_2)\sigma_z}(1+mA)\psi(x_1),
\end{align}
where
\begin{align}
A &= -\int dx' \{\cos\phi(x')\sigma_y-\sin\phi(x')\sigma_x\} \nonumber \\
&\equiv S\sigma_x+C\sigma_y.
\end{align}
The localization length $\lambda$ can be found by ($X\equiv \langle \sigma_x \rangle$, $Y\equiv \langle \sigma_y \rangle$)
\begin{align}
e^{-2(\frac{x_2-x_1}{\lambda})}&\sim 1-2(\frac{x_2-x_1}{\lambda}) = \langle(1+mA)^2\rangle \nonumber\\&= 1 + 2m(SX+CY)+m^2(C^2+S^2)
\label{eq:loc_len}
\end{align}
At $m =  0$, the uniform (Haar measure) distribution on the unit sphere is a 
stationary distribution. Therefore, this can be solved in the small $m$ limit.
For a stationary distribution, expectation value of a local operator $\hat{O}$ should satisfy
\begin{align}
\frac{\langle (1+mA)e^{i\phi(x_2)\sigma_z} \hat{O}e^{-i\phi(x_2)\sigma_z}(1+mA)\rangle}{\langle (1+mA)^2\rangle} = \langle\hat{O}\rangle.
\end{align}
By choosing $\hat{O}=\sigma_{x,y}$, we obtain the following equations, 
\begin{align}
X &=\frac{\langle(1+mA)[\cos\phi(x_2) \sigma_x -\sin\phi(x_2) \sigma_y](1+mA)\rangle}{\langle (1+mA)^2\rangle} \nonumber \\
Y &=\frac{\langle(1+mA)[\sin\phi(x_2) \sigma_x +\cos\phi(x_2) \sigma_y](1+mA)\rangle}{\langle (1+mA)^2\rangle}.
\end{align}
To lowest order in m,
\begin{align}
\begin{pmatrix} 
X  \\
Y  
\end{pmatrix}
&=\frac{m}{1-\cos\phi(x_2)}\nonumber\\
&\times
\begin{pmatrix} 
\cos\phi(x_2)-1 & -\sin\phi(x_2) \\
\sin\phi(x_2) & \cos\phi(x_2)-1
\end{pmatrix}
\begin{pmatrix} 
S  \\
C  
\end{pmatrix}.
\end{align}
Plugging X and Y into Eq.~\ref{eq:loc_len}, we find that
\begin{align}
2(\frac{x_2-x_1}{\lambda}) = m^2(C^2+S^2)
\end{align}
Next, we solve for the $C^2+S^2$. This involves averaging over disorder potential $\mu(x) = E + \xi(x)$, where $\langle\langle\xi(x)\xi(y)\rangle\rangle = V^2\Lambda \delta (x-y)$ is uncorrelated,
\begin{align}
C^2+S^2 &=\int_{x_1}^{x_2}\langle\langle\cos\phi(x-x')\rangle\rangle dxdx' \nonumber\\
&=2\int_{x_1}^{x_2}dx\int^x_{x_1}dx'\cos (E(x-x'))e^{-\frac{V^2\Lambda}{2}(x-x')}
\end{align}
In the large $x_2-x_1$ limit, this gives
\begin{align}
C^2+S^2 \sim &\frac{4V^2\Lambda}{4E^2+V^4\Lambda^2}(x_2-x_1).
\end{align}
Finally, the localization length is
\begin{align}
\lambda = \frac{4E^2+V^4\Lambda^2}{2m^2V^2\Lambda}
\end{align}

\section{Self-energy Evaluation of the JJC Model with QPS}
The microscopic Hamiltonian with quantum phase slip is 
\begin{align}
H = \sum_q \frac{\pi^2 E_J}{2}\Pi_q^2+\frac{4\sin^2(q/2)U_q}{\pi^2}\phi_q^2+g\sum_i \cos2(\bar{\phi}_i+\Lambda_i)
\end{align}
To write in path integral form, we normal-order the interaction term, resulting in an additional factor,
\begin{align}
\sum_i \cos2(\bar{\phi}_i+\Lambda_i) = \sum_{i}\xi_i :\cos2(\bar{\phi}_i+\Lambda_i):
\end{align}
where $\xi_i= \exp[-\frac{\pi^2}{8}\sum_q\sqrt{\frac{E_J}{2\sin ^2(q/2)U_q}} (A^i_q)^2]$ and $A^i_q$ is the momentum space coefficient of $\phi_q$ in $2\bar{\phi}_i = \sum_q A^i_q \phi_q$.
This leads to the action in terms of rescaled $\phi\rightarrow \phi' = \pi\sqrt{E_J}\phi$ 
\begin{align}
S[\phi] = &\sum_{q,\omega}\frac{1}{2}(\omega^2-\Omega_q^2)|\phi_{q,\omega}|^2 \nonumber\\ &- g\int dt\sum_i\xi_i\cos2(\pi\sqrt{E_J}\bar{\phi}_i+\Lambda_i)
\end{align}
at low energy, we can expand the interaction term in small $\bar{\phi}$. Next, since we are focusing on the self-energy correction to the correlation function, we consider the cubic term, which will give the lowest order contribution to the self-energy,
\begin{align}
&-\frac{g}{3!}\int dt\sum_i\xi_i\sin 2\Lambda_i(2\bar{\phi})^3_i\nonumber\\
&=-\frac{g}{3!}\sum_{\{q, \omega\}}2\pi\delta(\omega_1+\omega_2+\omega_3)\Gamma_{q_1,q_2,q_3}\phi_{q_1,\omega_1}\phi_{q_2,\omega_2}\phi_{q_3,\omega_3}
\end{align}
where
\begin{align}
\Gamma_{q_1,q_2,q_3} = (\pi\sqrt{E_J})^3\sum_i\xi_i \sin2\Lambda_i A^i_{q_1}A^i_{q_2}A^i_{q_3}.
\end{align}
This gives the action
\begin{align}
&S[\phi] = \sum_{q,\omega}\frac{1}{2}(\omega^2-\Omega_q^2)|\phi_{q,\omega}|^2 \nonumber\\ &- \frac{1}{3!}\sum_{\{q,\omega\}}\Gamma_{k_1,k_2,k_3} \phi_{q_1,\omega_1}\phi_{q_2,\omega_2}\phi_{q_3,\omega_3}2\pi\delta(\omega_1+\omega_2+\omega_3)
\end{align}
The diagrammatic expansion of this action gives rise to the following Feynman rules for propagators and vertices which is a straightforward generalization of the $\phi^4$ theory,
\begin{itemize}
  \item Each plasmon line corresponds to the propagator
  \begin{align}
  \frac{i}{\omega^2-\Omega_k^2+i\epsilon}
  \end{align}
  \item Each vertex corresponds to a factor of
	\begin{align}
	-\frac{ig}{3!}\Gamma_{k_1,k_2,k_3}2\pi\delta(\omega_1+\omega_2+\omega_3)
	\end{align}
\end{itemize}
Applying these rules gives us the following formula for the lowest order self-energy,
\begin{align}
&\Sigma(k,\omega) = \mathcal{S}g^2\sum_{q_1,q_2}(-\frac{i}{3!}\Gamma_{k,q_1,q_2})^2\nonumber\\ &\times\int\frac{d\omega_1}{2\pi} \frac{i}{\omega_1^2-\Omega_{q_1}^2+i\epsilon}\frac{i}{(\omega+\omega_1)^2-\Omega_{q_2}^2+i\epsilon}\nonumber \\
\end{align}
plugging in the symmetry factor $\mathcal{S} = (3\times 3\times 2)\times 2$ and rewriting the correlation functions, 
\begin{align}
&\Sigma(k,\omega)=\sum_{q_1,q_2}(\Gamma_{k,q_1,q_2})^2(\frac{1}{4\Omega_{q_1}\Omega_{q_2}})\nonumber\\ &\int\frac{d\omega_1}{2\pi} (\frac{1}{\omega_1 -\Omega_{q_1}+i\epsilon}-\frac{1}{\omega_1+\Omega_{q_1}-i\epsilon})\nonumber\\
&\qquad\times(\frac{1}{(\omega+\omega_1)-\Omega_{q_2}+i\epsilon}-\frac{1}{\omega+\omega_1+\Omega_{q_2}-i\epsilon})\nonumber\\
&= i\sum_{q_1,q_2}(\Gamma_{k,q_1,q_2})^2(\frac{1}{4\Omega_{q_1}\Omega_{q_2}})\nonumber\\&\qquad\times[\frac{1}{-\omega+\Omega_{q_1}+\Omega_{q_2}-2i\epsilon}-\frac{1}{\omega+\Omega_{q_1}+\Omega_{q_2}-2i\epsilon}]
\end{align}
This gives 
\begin{align}
G = \frac{1}{G_0^{-1}-\Sigma}=\frac{i}{\omega^2-\Omega_k^2-i\Sigma}
\end{align}
with the pole shifted to
\begin{align}
\Omega_k+i\frac{\Sigma(k,\Omega_k)}{2\Omega_k}
\end{align}
This gives the lifetime of the plasmon by
\begin{align}
\frac{1}{\tau_k} &= \mathbf{Im}[-i\frac{\Sigma(k,\Omega_k)}{2\Omega_k}]\nonumber\\
&=\pi\sum_{q_1,q_2}(\Gamma_{k,q_1,q_2})^2\frac{\delta(\Omega_k-\Omega_{q_1}-\Omega_{q_2})}{8\Omega_k\Omega_{q_1}\Omega_{q_2}}
\end{align}
where we have used the fact that all $\Omega$'s are positive, therefore $\delta(\Omega_k+\Omega_{q_1}+\Omega_{q_2})$ can not be satisfied.

\end{document}